\documentclass{nature}
\usepackage{graphicx}
\makeatletter
\let\saved@includegraphics\includegraphics
\AtBeginDocument{\let\includegraphics\saved@includegraphics}
\renewenvironment*{figure}{\@float{figure}}{\end@float}
\makeatother
\usepackage{amssymb}
\usepackage[breaklinks,colorlinks,urlcolor=blue,citecolor=blue,linkcolor=blue]{hyperref}
\usepackage{amsmath,amstext}
\usepackage[numbers,sort&compress]{natbib}
\newcommand       \tyc          {TYC 2597-735-1}
\newcommand       \meth	        {SI}	
\newcommand       \si     		  {SI}	

\newcommand{\unit}[1]{\ensuremath{\, \mathrm{#1}}}

\title{A blue ring nebula from a stellar merger several thousand years old}

\author{Keri Hoadley$^{1,10,12}$, D.~Christopher Martin$^{1,12}$, Brian D.~Metzger$^{2,3,12}$, Mark Seibert$^{4,12}$, Andrew McWilliam$^4$, Ken J.~Shen$^5$, James~D. Neill$^1$, Gudmundur Stefansson$^{6,7,8,11}$, Andrew Monson$^{7,8}$ \& Bradley E. Schaefer$^9$}

\begin{document}


\newcommand{\be}{\begin{equation}}
\newcommand{\ee}{\end{equation}}

\maketitle

\begin{affiliations}
 \item Cahill Center for Astrophysics, California Institute of Technology, 1216 East California Boulevard, Mail Code 278-17, Pasadena, California 91125, USA.
 \item Department of Physics, Columbia Astrophysics Laboratory, Columbia University, New York, NY 10027, USA.
 \item Center for Computational Astrophysics, Flatiron Institute, New York, NY 10010, USA
 \item The Observatories of the Carnegie Institution for Science, 813 Santa Barbara Street, Pasadena, CA 91101, USA.
 \item Department of Astronomy \& Theoretical Astrophysics Center, University of California at Berkeley, 501 Campbell Hall, Berkeley, CA 94720, USA.
 \item Department of Astrophysical Sciences, Princeton University, 4 Ivy Lane, Princeton, NJ 08540, USA
 \item Department of Astronomy \& Astrophysics, The Pennsylvania State University, 525 Davey Lab, University Park, PA 16802, USA
 \item Center for Exoplanets \& Habitable Worlds, University Park, PA 16802, USA
 \item Department of Physics \& Astronomy, Louisiana State University, Baton Rouge, LA 70820, USA.
 \item David \& Ellen Lee Caltech Prize Postdoctoral Fellow in Experimental Physics
 \item Henry Norris Russell Fellow
 \item Authors have contributed equally
\end{affiliations}

\begin{abstract}	

Stellar mergers are a brief-lived but common phase in the evolution of binary star systems\cite{Sana+12,Temmink+20}. Among the many astrophysical implications of these events include the creation of atypical stars (e.g. magnetic stars\cite{Schneider+19}, blue stragglers\cite{Davies+04}, rapid rotators\cite{Leiner+19}), interpretation of stellar populations\cite{Wang+2020}, and formation channels of LIGO-detected compact object mergers\cite{Belczynski+18}. Although stellar mergers are thus commonly invoked phenomena, observations of these events and details of their evolution remain elusive. While a  handful of stellar mergers have been directly observed in recent years\cite{Bond+03,Kulkarni+07}, the central remnants of these events remain shrouded by an opaque shell of dust and molecules\cite{Tylenda&Kaminski16}, making it impossible to observe their final state (e.g. as a single merged star or a tighter surviving binary\cite{Ivanova+13}). Here we report observations of an unusual, ring-shaped ultraviolet nebula and the star at its center, TYC 2597-735-1. The nebula shows two opposing fronts, suggesting a bipolar outflow from TYC 2597-735-1. TYC 2597-735-1’s spectrum and proximity above the Galactic plane suggest it is an old star, yet it shows abnormally low surface gravity and a detectable long-term luminosity decay, uncharacteristic for its evolutionary stage. TYC 2597-735-1 also exhibits H-alpha emission, radial velocity variations, enhanced ultraviolet radiation, and excess infrared emission, common signposts of dusty circumstellar disks\cite{Adams+87}, stellar activity\cite{Figueira+2013}, and accretion\cite{Lima+10}. The combined observations, paired with stellar evolution models, suggests TYC 2597-735-1 merged with a lower-mass companion several thousand years ago. TYC 2597-735-1 provides a look at an unobstructed stellar merger found at an evolutionary stage between its dynamic onset and the theorized final equilibrium state, directly inferring how two stars merge into a single star.   

\end{abstract}

The ``blue'' ring nebula (Figure~1(a)) is a rare far-ultraviolet emitting object discovered by the {\it Galaxy Evolution Explorer}\cite{Martin+2005}. The nebula has not been observed in any other part of the electromagnetic spectrum to date (Extended Data (ED) Figure~\ref{fig:brn_compilation}; see Supplementary Information (SI)). The nebula is ring-shaped and smooth, extending $\sim$8$^{\prime}$ across the sky at a slightly inclined (15 degrees), face-on view.  
Like other extended, far-ultraviolet sources\cite{Martin+07}, molecular hydrogen (H$_2$), which fluoresces throughout the far-ultraviolet ($\lambda$ $<$ 1700{\AA}; ED Figure~\ref{fig:fuv_grism}), is responsible for the nebular emission.  The total luminosity of the nebula, $\sim 3\times$10$^{33}$ erg s$^{-1}$ (see \meth), yields an upper limit of 10$^{44}$ H$_2$ molecules fluorescing per second. With a 15\% chance of destroying H$_2$ in the fluorescence process, 
the rate of H$_2$ destruction in the nebula is $\lesssim$2.5$\times$10$^{-14}$ solar masses (M$_{\odot}$) per second.  Combined with the age of the nebula (see below), we estimate a minimum nebular mass $M_{\rm BRN} \gtrsim 0.004$ M$_{\odot}$ (or $\sim$4 Jupiter masses; see \meth). 

Lining the western edge of the ultraviolet nebula appears a thin, faint shock filament seen in near-ultraviolet, far-ultraviolet, and H$\alpha$ emission (Figure~1(b,c,d)). Ground-based H$\alpha$ imaging shows the filament is part of a more extended shock system (Figure~1(d)), comprised of two overlapping, circular rings, which are offset by $\sim$3$^{\prime}$ on the sky (Figure~1(e,f)). We measure the velocity of the shock using Keck/LRIS multi-slit spectroscopy, finding the two circular rings expanding in opposite directions directions with v$_{\rm shock} \pm$ 400 kilometers per second.  Neutral hydrogen (H$\alpha$, H$\beta$, etc.) emits in these filaments, with line ratios suggesting formation in a non-radiative shock created as the outflow sweeps up and heats interstellar gas (see \meth).  

The large-scale geometry and inference of dual shock fronts expanding in opposite directions indicate a bipolar outflow originating from a star at the center of the nebula, \tyc~(Figure~1(f)). \tyc~is located 1.9 kiloparsecs away\cite{Gaia+18} at 1.5 kiloparsecs above the Galactic plane, suggesting the nebula extends $\sim$4 parsecs.  Pairing the physical extent of the nebula with its velocity, we limit the age of the nebula to $<$5,000 years (see \meth).  

Spectral synthesis models of \tyc's Keck/HIRES optical spectra find its stellar luminosity $L_{\star} \simeq 110L_{\odot}$, effective temperature $T_{eff} \approx 5850$ K, surface gravity $g \approx 600$ cm/s$^{2}$, and stellar radius $R_{\star} \approx 11R_{\odot}$ (see \meth). These spectroscopic parameters suggest \tyc~has a mass around 1 -- 2.1 M$_{\odot}$ and has evolved off the main sequence.  Notably, \tyc~appears puffier for its temperature than other evolved stars of similar luminosity (ED Figure~\ref{fig:tefflogg}).
  
The surface iron abundance of \tyc~is sub-solar ($[{\rm Fe/H}] \sim -0.9$ dex), while its $\alpha$-element abundances appear enhanced ($[\alpha/{\rm Fe}] \sim +0.4$ dex; see \meth). Its parallax, proper motions, and radial velocity indicate disk-like kinematics (UVW $\sim$ 7.8, 11.3, 36.0 km/s), based on solar peculiar motions. These observations support \tyc's membership in the thick-disk population\cite{Ness+13}.

\tyc~also displays prominent H$\alpha$ line emission, excess far-ultraviolet flux, and radial velocity variations.  The H$\alpha$ emission varies in both line shape and amplitude on timescales of days, showing an enhanced blue-shifted component (ED Figure~\ref{fig:halpha_bisector}). The observed far-ultraviolet magnitude of \tyc~is over 6 orders of magnitude brighter than is expected from synthetic stellar models (Figure~2(a)). Radial velocity measurements of \tyc~find $\sim$200 meters per second Doppler shift, yet exclude the presence of a binary companion with mass $\gtrsim 0.01 M_{\odot}$ in tight orbit around \tyc~($a \lesssim$ 0.1 astronomical units; ED Figure~\ref{fig:rv}). Altogether, these signatures point to heightened stellar surface activity at \tyc.  

Additionally, \tyc~emits excess infrared radiation (Figure~2(a); see \meth), a tell-tale sign of a dusty circumstellar disk\cite{Adams+87}.  A disk-like geometry, which lies in a plane perpendicular to our line of sight and the symmetric axis of the nebula, is favored because it lacks evidence of circumstellar reddening, as measured from \tyc's optical spectrum (see \meth).  The combination of infrared emission and surface activity paint a picture where \tyc~is actively accreting material from a disk of gas and dust extending to several astronomical units (see \meth).  Other systems with similar observable traits (e.g., T Tauri stars\cite{Edwards+1987,Lima+10} and AGB stars with accretion disks\cite{Sahai+2008}), are often interpreted as actively accreting material from circumstellar disks.

Any viable explanation for \tyc~and its ultraviolet nebula should account for the unusual properties of the star, its circumstellar environment, and the fast outflow launched from its vicinity $<$5,000 years ago.  Protostellar systems exhibit many of the same characteristics as \tyc, including H$\alpha$ emission and excess ultraviolet flux\cite{Fukui+1989}; however, its isolation from star-forming environments and proximity above the Galactic plane strongly disfavor it being a young protostar. At the other end of the stellar evolution spectrum, \tyc's stellar properties do not match those of ``post-red giant" systems, which are more luminous ($>$1,000 L$_{\odot}$). While evolved stars are known to host dusty disks and expel massive stellar winds\cite{Kamath+16}, the velocity of the ultraviolet nebula (v$_{\rm shock}$ $\sim$ 400 km/s) is $\gtrsim 10$ times faster than those measured from (post-)asymptotic giant branch stars and pre-planetary nebulae (v$_{\rm wind}$ $\sim$ 10's km/s)\cite{Bujarrabal+2018}. Radial velocity measurements exclude a short-period companion orbiting \tyc~capable of ejecting a collimated, bipolar outflow with its observed velocity (see \meth), thus disfavoring classical novae, cataclysmic variability, or other mass-transfer interactions with a surviving compact object.

The mass ejection event responsible for this ultraviolet nebula and the unusual present state of \tyc~appear most consistent with a binary star merger.  To test this scenario and estimate the initial state of the system, we use the stellar evolution code MESA\cite{Paxton+19} to explore the impact a stellar merger has on long-term stellar properties\cite{Metzger+17} (see \meth). We find that a low-mass companion ($M_{\rm c} \sim 0.1 M_{\odot}$) reasonably reproduces \tyc's effective temperature, luminosity, and surface gravity at a post-merger age of $t_{\rm age} \approx 1,000$ years, accounting for \tyc's position in $T_{eff}$-log$g$ space (ED Figure~\ref{fig:tefflogg}). These models also predict that \tyc~was $\approx 0.1$ B-mag brighter a century ago, which we observe in historical DASCH archive records of \tyc~(see \meth, ED Figure~\ref{fig:lc}).
 
The best-fit models are those where the merger happens after the primary begins evolving off its main sequence (ED Figure~\ref{fig:mesa}). Such timing may not be coincidental if the companion was dragged into the star through tidal interaction (Figure~3(a)): the timescale for tidal orbital decay depends sensitively on the primary radius ($\tau_{tide} \sim R_{\star}^{-5}$)\cite{Goldreich+1966}, accelerating the orbital decay as the primary reaches its sub-giant phase.  Indeed, tidally induced mergers may explain the dearth of short-period planets around evolved A-stars\cite{Johnson+07}.  

The events leading up to and following the merger of \tyc~and its companion shape the system we see today.  As \tyc~and its companion approached sufficiently closely, the former overflowed its Roche lobe onto the latter, initiating the merger (Figure~3(b)).  Numerical simulations demonstrate that the earliest phase of the merger process results in the ejection of matter through the outer $L_{2}$ Lagrange point in the equatorial plane of the binary\cite{Pejcha+16a}.  Most of this matter remains gravitationally bound around the star system, forming a circumbinary disk.  The companion, unable to accommodate the additional mass, is dragged deeper into the primary's envelope in a runaway process\cite{Ivanova+13} (Figure~3(c)).  This delayed dynamical phase is accompanied by the ejection of a shell of gas. A portion of the ejected material is collimated by the circumbinary disk into a bipolar outflow\cite{MacLeod+18}. The balance of the mass lost during primary-companion interactions remains as a circumstellar disk, which spreads out and cools over time, eventually reaching sufficiently low temperatures to form dust.  

We see evidence of this relic disk around \tyc~today as an infrared excess. A simple analytic model, which follows the spreading evolution of the gaseous disk due to internal viscosity over the thousands of years since the merger (see \si), is broadly consistent with both the present-day gas accretion rate (estimated from H$\alpha$; see \meth) and a lower limit on the present-day disk mass obtained by fitting the infrared spectral energy distribution of \tyc~($M_{\rm disk,dust} \gtrsim 5\times 10^{-9}  M_{\odot}$; see \meth). Accretion of disk material onto \tyc~could account for its observed stellar activity (e.g., H$\alpha$ emission and far-ultraviolet excess).  Angular momentum added to the envelope of the star by the merger, and subsequent accretion of the disk, would also increase \tyc's surface rotation velocity. Indeed, we find the de-projected surface rotation velocity of \tyc~to be $\approx 25$ km/s (see \meth, ED Figure~\ref{fig:rot_velocity}), larger than expected for a star which has just evolved off the main sequence ($v <$ 10 km/s)\cite{deMedeiros+1996}.  

The bipolar ejecta shell expands away from the stellar merger, cooling and satisfying within weeks the conditions for molecular formation and solid condensation (Figure~3(d)), as seen directly in the observed ejecta of luminous red novae\cite{Kaminski+18}.  The mass ejected, inferred from merger simulations and modeling the light curves of luminous red novae, is typically $\approx$ 0.01-0.1 $M_{\odot}$\cite{Pejcha+17}, consistent with our lower mass limit of the ultraviolet nebula (0.004 $M_{\odot}$). 

As the nebula expands and sweeps up interstellar gas, a reverse shock crosses through the ejecta shell, heating electrons in its wake. These electrons excite the H$_2$ formed in the outflow, which fluoresces in the far-ultraviolet (Figure~3(e)).  Although dust is almost always observed in the ejecta of stellar mergers\cite{Kaminski+18}, we find no evidence of dust in the ultraviolet nebula (e.g., ultraviolet/optical reddening).  We speculate that either the dust was destroyed in the reverse shock\cite{Martinez-Gonzalez+2019} or that the bipolar ejecta shell has thinned out sufficiently such that dust is currently undetectable.  Consistent with the latter, this ultraviolet nebula marks the oldest observed stellar merger to date, being $\gtrsim 3-10$ times older than the previously oldest stellar merger candidate, CK Vulpeculae (1670), which is still shrouded by dust\cite{Kaminski+17}. The system was caught at an opportune time - old enough to reveal the central remnant, yet young enough that the merger-generated nebula has not dissolved into the interstellar medium.

The discovery of an ultraviolet nebula introduces a new way of identifying otherwise hidden late-stage stellar mergers.  With 1 -- 10 of these objects expected to be observable in the Milky Way (see \meth), future far-ultraviolet telescopes may uncover more late-stage stellar mergers. \tyc~poses a unique opportunity to study post-merger morphology as the only known merger system not enshrouded by dust. For example, the close separation of the initial stellar binary$-$which shares properties broadly similar to those of protoplanetary disks$-$could form ``second generation'' planets\cite{Schleicher+14}.  However, given the relatively short time ($\lesssim 10^{8}$ years) until \tyc~reaches the end of its nuclear burning life, the potential window of habitability for such planets could be drastically reduced compared to main sequence stars.


\begin{addendum}
	\item[Supplementary Information] is linked to the online version of the paper at www.nature.com/nature.

	\item This research is based on observations made with the \emph{Galaxy Evolution Explorer}, obtained from the MAST data archive at the Space Telescope Science Institute, which is operated by the Association of Universities for Research in Astronomy, Inc., under NASA contract NAS 5–26555. 
Some of the data presented herein were obtained at the W. M. Keck Observatory, which is operated as a scientific partnership among the California Institute of Technology, the University of California and the National Aeronautics and Space Administration. This research has made use of the Keck Observatory Archive (KOA), which is operated by the W. M. Keck Observatory and the NASA Exoplanet Science Institute (NExScI), under contract with the National Aeronautics and Space Administration. The Observatory was made possible by the generous financial support of the W. M. Keck Foundation. The authors wish to recognize and acknowledge the very significant cultural role and reverence that the summit of Maunakea has always had within the indigenous Hawaiian community. We are most fortunate to have the opportunity to conduct observations from this mountain. 
Some of the data presented herein were obtained at the Palomar Observatory. 
This research has made use of the NASA/IPAC Infrared Science Archive, which is operated by the Jet Propulsion Laboratory, California Institute of Technology, under contract with the National Aeronautics and Space Administration. We especially thank Vicky Scowcroft for obtaining \emph{Spitzer}/IRAC photometry of \tyc. 
Funding for APASS has been provided by the Robert Martin Ayers Sciences Fund. The DASCH data from the Harvard archival plates was partially supported from National Science Foundation grants AST-0407380, AST-0909073, and AST-1313370. The American Association of Variable Star Observers has been critically helpful for finder charts, comparison star magnitudes, and recruiting skilled observers, including Sjoerd Dufoer, Kenneth Menzies, Richard Sabo, Geoffrey Stone, Ray Tomlin, and Gary Walker. 
These results are based on observations obtained with the Habitable-zone Planet Finder Spectrograph on the Hobby-Eberly Telescope. These data were obtained during HPF's Engineering and Commissioning period. We thank the Resident astronomers and Telescope Operators at the HET for the skillful execution of our observations of our observations with HPF. The authors would like to thank Caleb Ca$\rm\tilde{n}$as for providing an independent verification of the HPF SERVAL pipeline using a CCF-based method to calculate the RVs, which resulted in fully-consistent RVs to the SERVAL-based RVs presented here. The Hobby-Eberly Telescope is a joint project of the University of Texas at Austin, the Pennsylvania State University, Ludwig-Maximilians-Universität München, and Georg-August Universität Gottingen. The HET is named in honor of its principal benefactors, William P. Hobby and Robert E. Eberly. The HET collaboration acknowledges the support and resources from the Texas Advanced Computing Center. This work was partially supported by funding from the Center for Exoplanets and Habitable Worlds. The Center for Exoplanets and Habitable Worlds is supported by the Pennsylvania State University, the Eberly College of Science, and the Pennsylvania Space Grant Consortium.
This research made use of \texttt{photutils} and \texttt{astropy}, community-developed core \texttt{Python} packages for Astronomy, and Modules for Experiments in Stellar Astrophysics (MESA). 
The authors thank Prof. Armando Gil de Paz for obtaining the narrow-band filter H$\alpha$ imagery; Prof. John Johnson for commissioning \tyc~RV measurements as part of the California Planet Finder (CPF) program; and Prof. Andrew Howard for spearheading Keck/HIRES spectra and performing the primary RV reduction on all HIRES data. 
KH thanks Lynne Hillenbrand and Erika Hamden for productive discussions regarding critical aspects of this work. BDM acknowledges support from the Hubble Space Telescope ($\#$ HST-AR-15041.001-A) and from the National Science Foundation ($\#$ 80NSSC18K1708).  KJS received support from the NASA Astrophysics Theory Program (NNX17AG28G). GS and A.~Monson acknowledge support from NSF grants AST-1006676, AST-1126413, AST-1310885, AST-1517592, AST-1310875, AST-1907622, the NASA Astrobiology Institute (NAI; NNA09DA76A), and PSARC in our pursuit of precision radial velocities in the NIR with HPF. We acknowledge support from the Heising-Simons Foundation via grant 2017-0494 and 2019-1177. Computations for this research were performed on the Pennsylvania State University’s Institute for Computational \& Data Sciences (ICDS). GS acknowledges support by NASA HQ under the NASA Earth and Space Science Fellowship (NESSF) Program through grant NNX16AO28H. 

	\item[Author Contributions] Authors K.~Hoadley and B.~D.~Metzger organized and wrote the main body of the paper. Authors K.~Hoadley and M.~Seibert performed the data reduction and analysis of the \emph{GALEX} data, investigated the source of the ultraviolet emission, quantified the mass the far-ultraviolet nebula, and led the the analysis of \tyc's H$\alpha$ emission and variability. Author B.~D.~Metzger lead all theoretical and analytic interpretation efforts of the ultraviolet nebula origins and \tyc~in the context of stellar mergers and present-day luminous red novae.  Authors D.~C.~Martin and M.~Seibert spearheaded the \emph{GALEX} program that led to the detection of the ultraviolet nebula in 2004 and all subsequent follow-up observations of the nebula with {\it GALEX}. Both contributed to the overall interpretation of the observational data. Author D.~C.~Martin contributed to the organization and manifestation of the manuscript. Author M.~Seibert lead the radial velocity analysis, the interpretation and analysis of the infrared excess in \tyc's spectral energy distribution, and modeled the \tyc~spectral energy distribution (both stellar and dust infrared excess components). M.~Seibert coordinated all ground-based observations of the blue ring nebula and \tyc~at Palomar Observatory and W.~M.~Keck Observatory.  

Author A.~McWilliam derived the physical parameters and performed the model atmosphere chemical abundance analysis of \tyc. A.~McWilliam participated in the discussion of observations, analysis, and interpretation that led to the manifestation of this work. 
Author K.~J.~Shen performed the MESA calculations and participated in the discussion of observations, analysis, and interpretation that led to the manifestation of this work. 

Author J.~D.~Neill handled the data analysis and reported the subsequent result of the velocity structure of the H$\alpha$ shock observed with Keck/LRIS. J.~D.~Neill participated in the discussion of observations, analysis, and interpretation that led to the manifestation of this work.  

Author G.~Stefansson performed the HET/HPF radial velocity and differential line width indicator extractions and provided expertise on the interpretation of the combined radial velocity data sets.  

Author A.~Monson coordinated HET/HPF observations and both performed and reduced all TMMT B-band observations. 

Author B.~E.~Schaefer extracted and analyzed the long-term light curve data from 1897.5–2019.9.

	\item[Competing Interests] The authors declare that they have no competing financial interests.

	\item[Correspondence] Correspondence and requests for materials should be addressed to Dr.~Keri Hoadley (email: khoadley@caltech.edu).
	
	\item[Data availability] All {\it GALEX} imaging and grism data of \tyc~and its ultraviolet nebula are publicly available for download from the Mikulski Archive for Space Telescopes (MAST) in raw and reduced formats. The archive can be accessed either at \url{http://galex.stsci.edu/GalexView/} or \url{https://mast.stsci.edu/portal/Mashup/Clients/Mast/Portal.html}. All Keck-LRIS and Keck-HIRES data for \tyc~are publicly available for download from the Keck Observatory Archive (KOA). KOA can be accessed at \url{https://koa.ipac.caltech.edu/cgi-bin/KOA/nph-KOAlogin}. \tyc~raw photometric lightcurve frames, plates, and lightcurves from 1895 -- 1985 are publicly available and downloaded as a part of the DASCH: Digital Access to a Sky Century at Harvard program. The DASCH data base can be accessed at \url{https://projects.iq.harvard.edu/dasch}.  More recent photmetry for the lightcurve construction were obtained by independent observers and the American Association of Variable Star Observers (AAVSO). These data have been uploaded to a publicly available repository; please contact Corresponding Author for a link to the repository. All other photometric data for \tyc~was obtained from publicly archived ground- and space-based imaging and surveys, stored on the SIMBAD Astronomical Database (\url{http://simbad.u-strasbg.fr/simbad/}) and the NASA/IPAC Infrared Science Archive (\url{https://irsa.ipac.caltech.edu/frontpage/}).  The Hobby-Eberly Telescope does not have a publicly-available archive to access Habitable Planet Finder (HPF) data. We provide spreadsheets with the relevant data products from the HPF campaign for \tyc~in a publicly available repository: \url{https://github.com/oglebee-chessqueen/BlueRingNebula.git} 

\item[Code availability] We used and Modules for Experiments in Stellar Astrophysics (MESA)\cite{Paxton+19} for a portion of our analysis. While MESA is readily available for public use, we use a custom sub-routine and MESA inline code to produce the \tyc~merger evolution model presented in this work. We provide both in a public repository: \url{https://github.com/oglebee-chessqueen/BlueRingNebula.git}. 
Use the ATLAS9 pre-set grid of synthetic stellar spectra\cite{CastelliKurucz2003} to fit the \tyc~spectral energy distribution to representative stellar spectra. All synthetic stellar spectra are publicly available: \url{https://www.stsci.edu/hst/instrumentation/reference-data-for-calibration-and-tools/astronomical-catalogs/castelli-and-kurucz-atlas}. Portions of our analysis used Python packages \texttt{photutils}\cite{photutils+16} and \texttt{astropy}\cite{astropy+18}.
\end{addendum}


\clearpage

\begin{figure}
\centering
\includegraphics[width=1.0\textwidth]{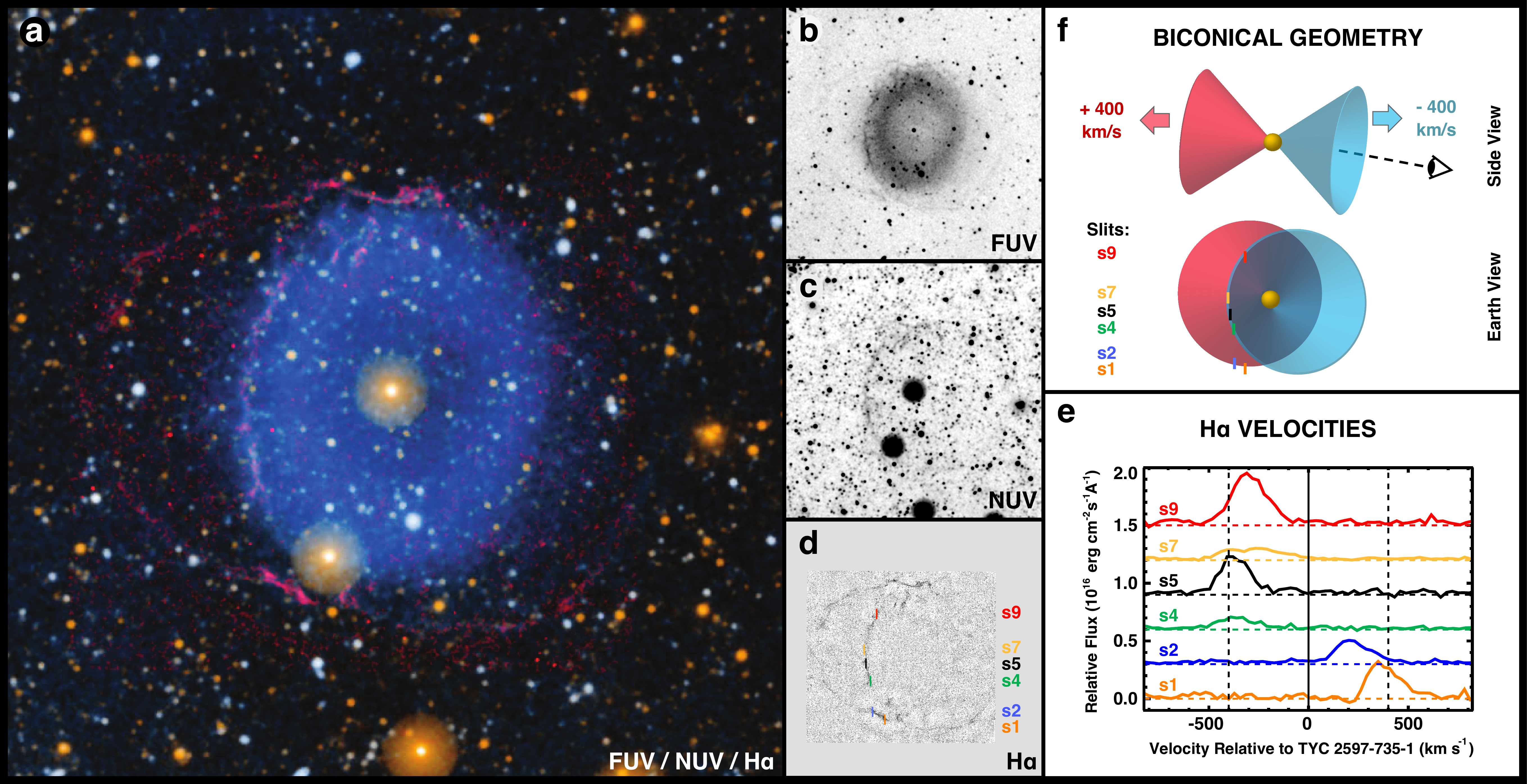} 
{\linespread{1.0}
\caption{
{\bf Ultraviolet and H$\alpha$ images of the blue ring nebula and a geometric schematic of the biconical outflow.} 
a. Three false-color images overlaid to show the nebula, with blue (far-ultraviolet: 1350 -- 1750 {\AA}), yellow (near-ultraviolet: 1750 -- 2800 {\AA}), and red (H$\alpha$). Field of view covers 15'$\times$15' on the sky. 
b. Far-ultraviolet image, showing emission in the central nebular region as well as in the elliptical shock filaments (fainter outskirts of the darker nebular region). 
c. Near-ultraviolet image, showing emission only in the shock filaments. 
d. H$\alpha$ image, showing emission only in the shock filaments. Slits are shown to demonstrate where along the shock filaments the velocity was measured. 
e. H$\alpha$ shock velocities along the back (red shifted) and front (blue shifted) cones. The slits overlaid on the shock filament are color-coded in panel a and f. Slits 1 and 2 (orange, blue) are on the back cone, the remainder on the forward cone.  
f. Biconical outflow model, side view and earth view with cone axis inclination 15 degrees (not to scale) and H$\alpha$ slits. The outflow velocity is $\pm$400 km/s. Emission in the far-ultraviolet comes from H$_2$ fluorescence and is only strong where the cones overlap (purple). The shocks emitting in the ultraviolet and H$\alpha$ only appear at the terminal periphery of the two cones. }
}	
\label{fig:brn}
\end{figure}

\begin{figure}
\centering
\includegraphics[width=0.75\textwidth]{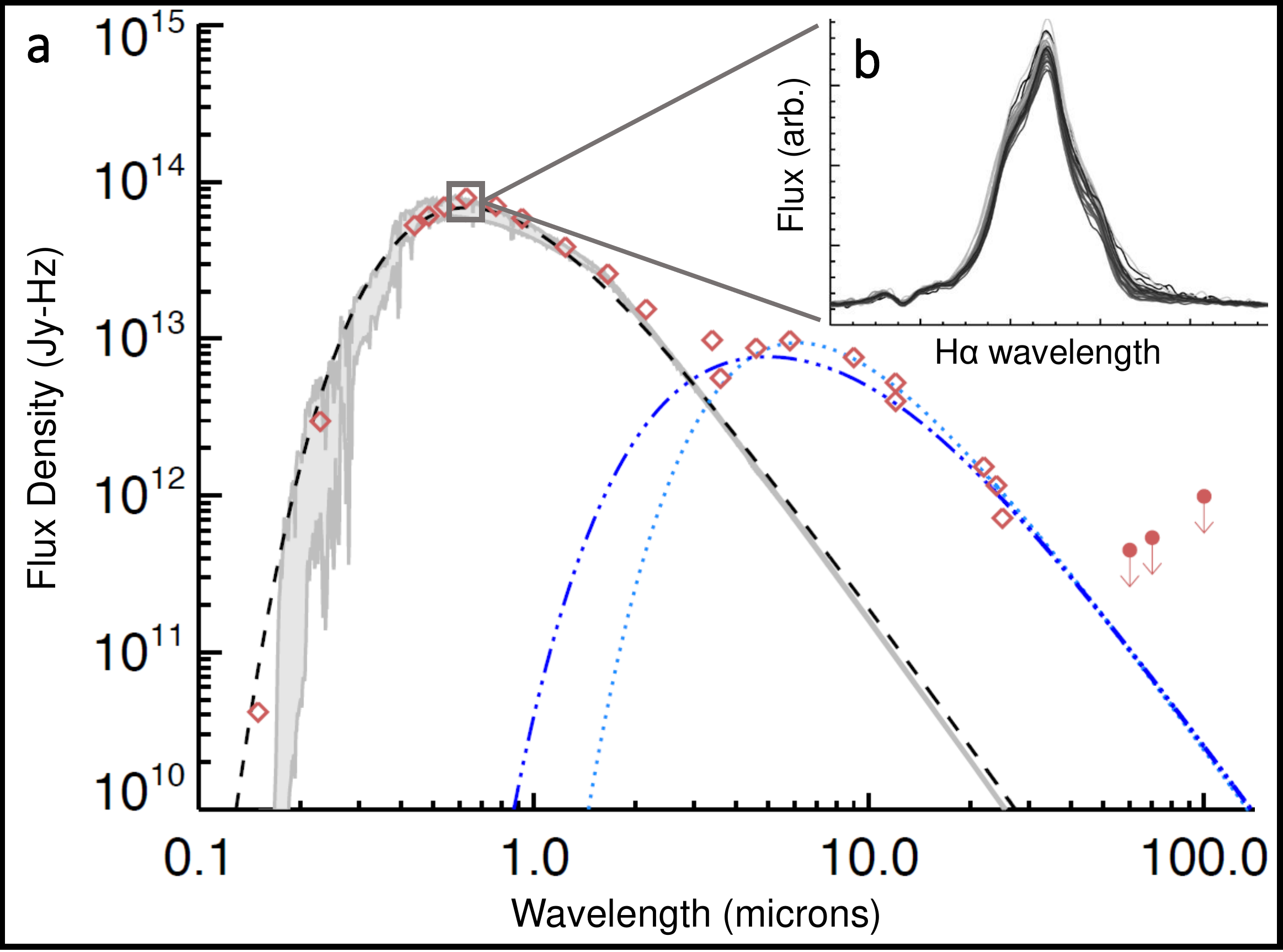} 
\vspace{0.5cm}
{\linespread{1.0}
\caption{
{\bf The spectral energy distribution and H$\alpha$ emission of \tyc.} 
a. The spectrum of TYC 2597-735-1, expressed as a function of spectral flux (in units of Jansky microns ($\mu$m)), shows enhanced near- and mid-infrared emission. Far-infrared measurements ($>$24 $\mu$m, \emph{IRAS}; solid circles) provide upper limits only. Synthetic stellar models that best match the inferred stellar properties of \tyc~(gray spectra and grayed region) do not account for either the infrared excess and also suggest a far-ultraviolet excess is present. Both the far-ultraviolet and infrared are frequently observed in systems actively accreting matter from warm, gaseous and dusty disks, like T Tauri protostars\cite{Adams+87}. Models of warm, dusty circumstellar disks reproduce the observed infrared excess (light blue dotted line: T$_{\rm dust} \sim$600 Kelvin, $\sim$0.2 - 3 astronomical units, assuming a disk inclination angle $\sim $15 degrees; dark blue dashed line: T$_{\rm dust} \sim$1200 - 300 Kelvin, $\sim$0.2 - 1.5 astronomical units, assuming a disk inclination angle $\sim $15 degrees).  
b. \tyc~exhibits H$\alpha$ \emph{emission}, an unusual trait for evolved stars. The H$\alpha$ line profile shows variability over short timescales. There is an enhanced blue edge to the emission, a classic signature of infalling material and traditionally interpreted as accretion flows or disk winds\cite{Lima+10}. The H$\alpha$ emission, excess infrared emission, enhanced far-ultraviolet radiation, and radial velocity variations (see ED Figure~\ref{fig:rv}) all suggest \tyc~is actively accreting materiel from the disk creating the observed infrared excess emission. }
}
\label{fig:BRN_morph}
\end{figure}

\begin{figure}
\centering
\includegraphics[width=1.0\textwidth]{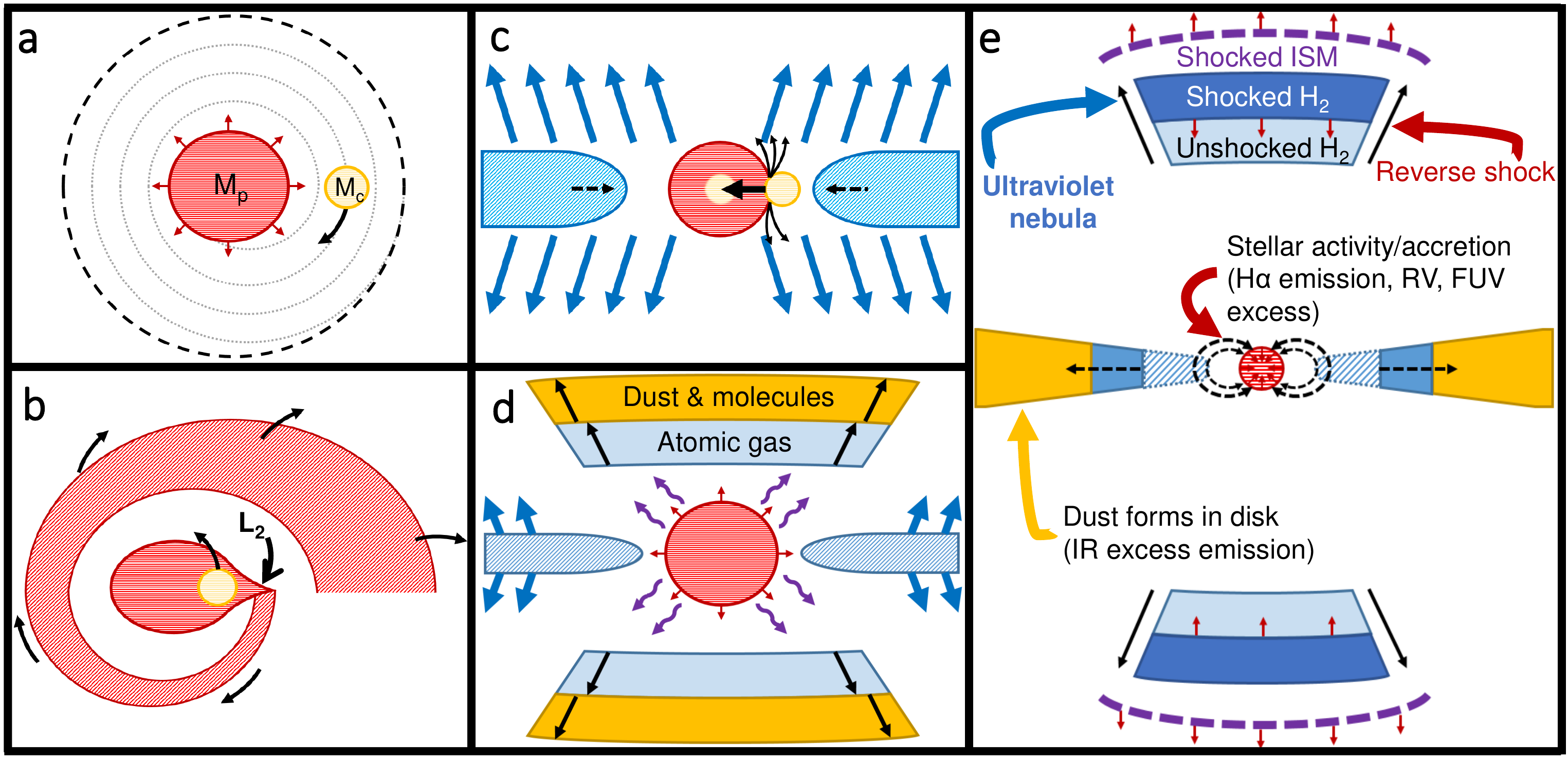} 
{\linespread{1.0}
\caption{
{\bf A schematic of the merger events responsible for the current state of TYC 2597-735-1 and its ultraviolet nebula (not to scale).} 
a. (Top view) As the primary star (M$_{p}$; red) evolves off the main sequence and its envelope expands, its companion (M$_{c}$; yellow) is slowly dragged inwards. 
b. (Top view) Over the course of many orbits, M$_{p}$ overflows its Roche lobe and deposits mass onto its companion. M$_{c}$, unable to hold on to this excess mass, spills it over into the common Lagrange point ($L_2$). M$_{c}$ begins to spiral into M$_{p}$. 
c. (Side view) M$_{c}$ plunges into M$_{p}$. Additional mass is ejected, shaped by the circumstellar disk formed by the $L_2$ overflow. 
d. (Side view) A bipolar outflow, ejected at speeds greater than or equal to the escape velocity of the system, expands and adiabatically cools, quickly forming dust and molecules. M$_{p}$ puffs up and brightens from the excess energy it received by consuming M$_{c}$. 
e. (Side view) Over the next thousand years, TYC 2597-735-1 slowly settles back to its equilibrium state. TYC 2597-735-1 displays activity, attributed to accretion flows fed by its remnant circumstellar disk (H$\alpha$ emission, RV, far-ultraviolet excess emission). The ejected outflow sweeps up interstellar material, initiating a reverse shock that clears the dust and excites H$_2$. The forward shock is seen in ultraviolet and H-alpha emission outlining the nebula today, while the reverse shock is revealed by the far-ultraviolet glow of H$_2$ fluorescence. }
}
\label{fig:brn_origins}
\end{figure}

\clearpage
\newpage

\renewcommand{\figurename}{Extended Data Figure}
\setcounter{figure}{0}

\begin{figure}
\centering
\includegraphics[angle=0, width=\textwidth]{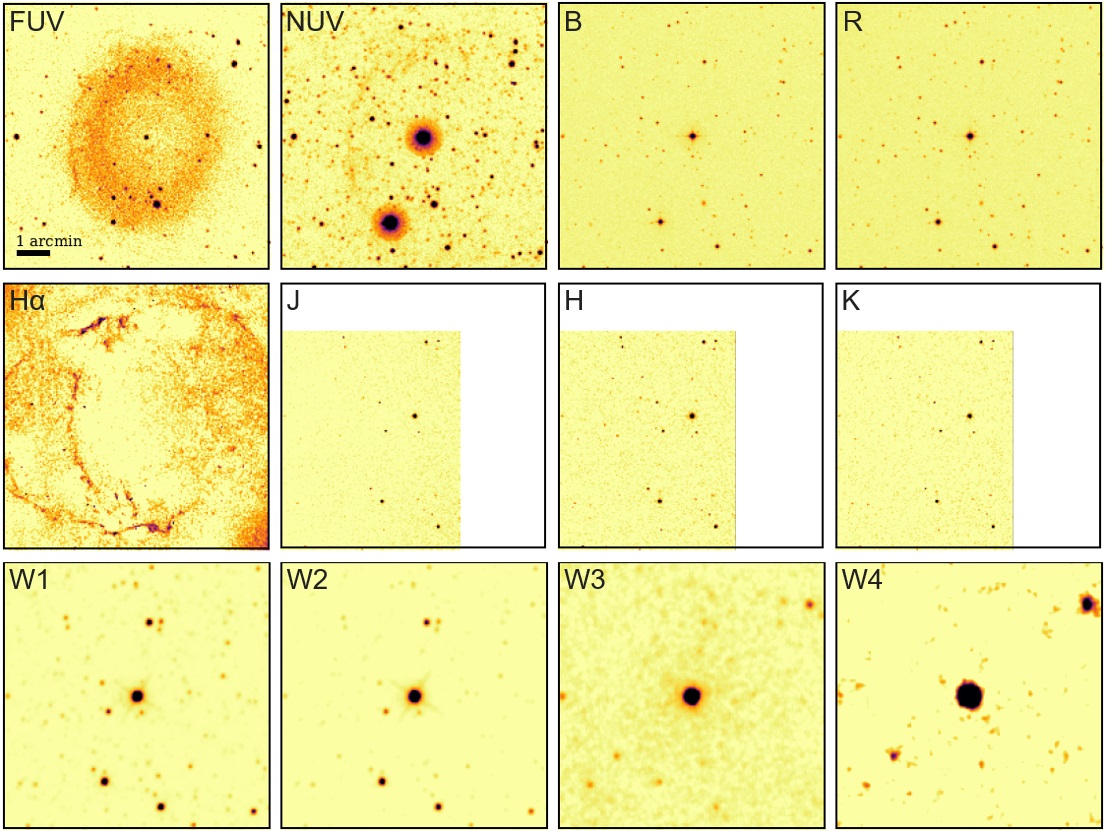}
{\linespread{1.0}
\caption{{\bf TYC 2597-735-1 and its ultraviolet nebula in different colors.} From left to right, top to bottom: \emph{GALEX} FUV, \emph{GALEX} NUV, DSS-II B-band, DSS-II R-band, Palomar Hale 200-inch COSMIC H$\alpha$ narrow-band, 2MASS J-band, 2MASS H-band, 2MASS K-band, and \emph{WISE} 3.4 (W1), 4.6 (W2), 12 (W3), and 22 (W4) $\mu$m. A reference line for 1 arcminute is included in the \emph{GALEX} FUV image. At a distance of 1.93 kpc, 1 arcminute corresponds to 0.56 pc. Each images covers 10$^{\prime} \times$10$^{\prime}$. All images are scaled by asinh to accentuate any faint, diffuse emission. The \emph{GALEX} NUV has been scaled to show the BRN western shock, which makes the brighter NUV stars (including \tyc) more enhanced.}
\label{fig:brn_compilation}}
\end{figure}

\begin{figure}
\centering
\includegraphics[angle=0, width=0.6\textwidth]{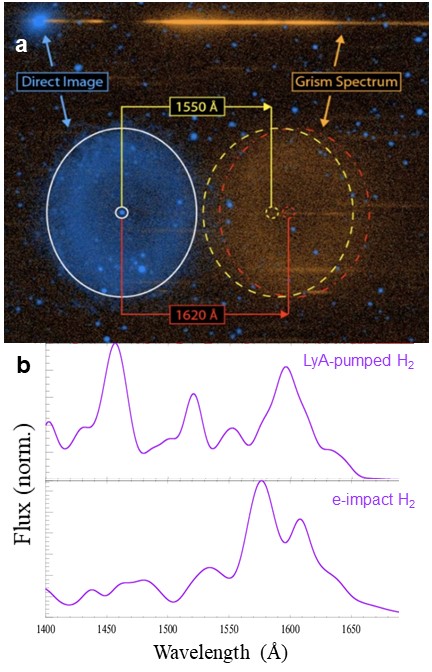}
{\linespread{1.0}
\caption{{\bf The source of emission in the far-ultraviolet nebula.} (Top) \emph{GALEX} low-resolution FUV grism imaging reveals the BRN only emits light around 1600{\AA}. Together with the lack of NUV radiation in the BRN, this points to  H$_2$ fluorescence as the main source of the BRN emission. (Bottom) Synthetic models of H$_2$ fluorescence change the distribution of light produced by H$_2$ in the FUV, depending on the source of excitation (examples show are Ly$\alpha$ photon pumping (top spectrum) and electron-collisional excitation (bottom spectrum). We convolved high-resolution synthetic H$_2$ fluorescence spectra with the \emph{GALEX} grism spectral resolution to produce the plots shown. Pumping by Ly$\alpha$ photons (whose source would likely come directly from \tyc) creates peaks in the distribution that are not seen by \emph{GALEX} near 1450{\AA}. Electron-impact fluorescence produces a spectral distribution that better matches where \emph{GALEX} sees the FUV emission being produced.}
\label{fig:fuv_grism}}
\end{figure}

\begin{figure}
\centering
\includegraphics[angle=0, width=0.75\textwidth]{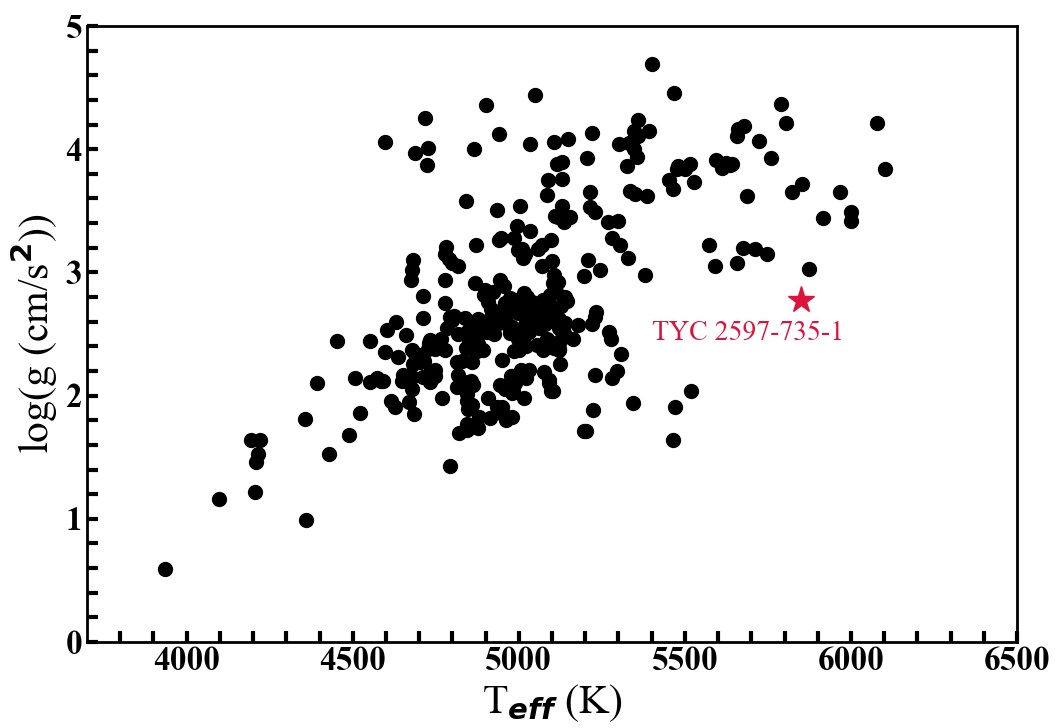}
{\linespread{1.0}
\caption{{\bf \tyc~is an outlier when compared with other moderately-evolved stars of similar mass.} A large sample of moderately evolved stars is shown\cite{Afsar+18} to demonstrate that, in particular, \tyc's effective temperature and surface gravity are not consistent with the majority of other stars following similar evolutionary tracks. If the present-day observable properties of \tyc~are a consequence of a previous stellar merger, as our MESA models suggest, then we expect that \tyc~is currently puffed up more than usual and will continue to relax over the next thousands of year to better match the trend of evolving stars in T$_{\rm eff}$-log(g) space (see ED Figure~\ref{fig:mesa}).}
 \label{fig:tefflogg}}
\end{figure}

\begin{figure} 
\centering
\includegraphics[angle=0, width=0.75\textwidth]{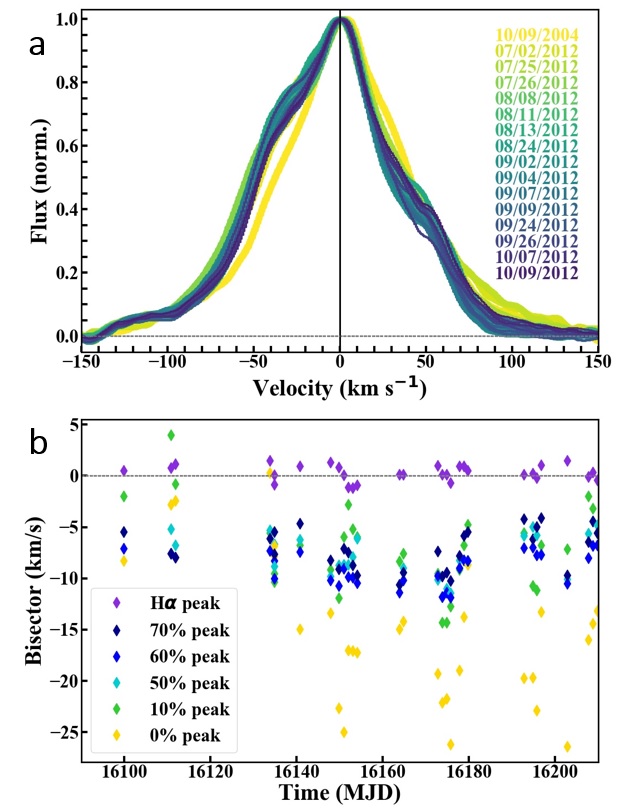}
{\linespread{1.0}
\caption{{\bf Stellar H$\alpha$ emission properties of \tyc.} a. \tyc~exhibits H$\alpha$ emission, an unusual trait for evolved stars. The H$\alpha$ line profile shows variability over short timescales. There is an enhanced blue edge to the emission, a traditional signature supporting gaseous accretion or disk winds\cite{Lima+10}. The H$\alpha$ emission suggests \tyc~is actively accreting matter, possibly from the disk creating its observed infrared excess emission. b. The H$\alpha$ bisector velocities at different parts of the H$\alpha$ emission line profile as a function of time (day). Different color diamonds (labeled in the figure) represent different flux levels in the line profile probed to determine the bisector value. The points in the line profile plotted show the most dramatic shifts away from the line center. The line peak bisector is also shown (purple), to demonstrate the day-by-day variability observed in the line profile. A dashed gray line represents no velocity shift from H$\alpha$ wavelength center. Except for the line peak, which fluctuates around $\sim$0 km/s shifts, the rest of the line profile tends towards negative bisector velocity values, providing evidence that the line profile tends towards a blue-shifted enhancement.}
\label{fig:halpha_bisector}}
\end{figure}

\begin{figure}
\centering
\includegraphics[angle=0, width=\textwidth]{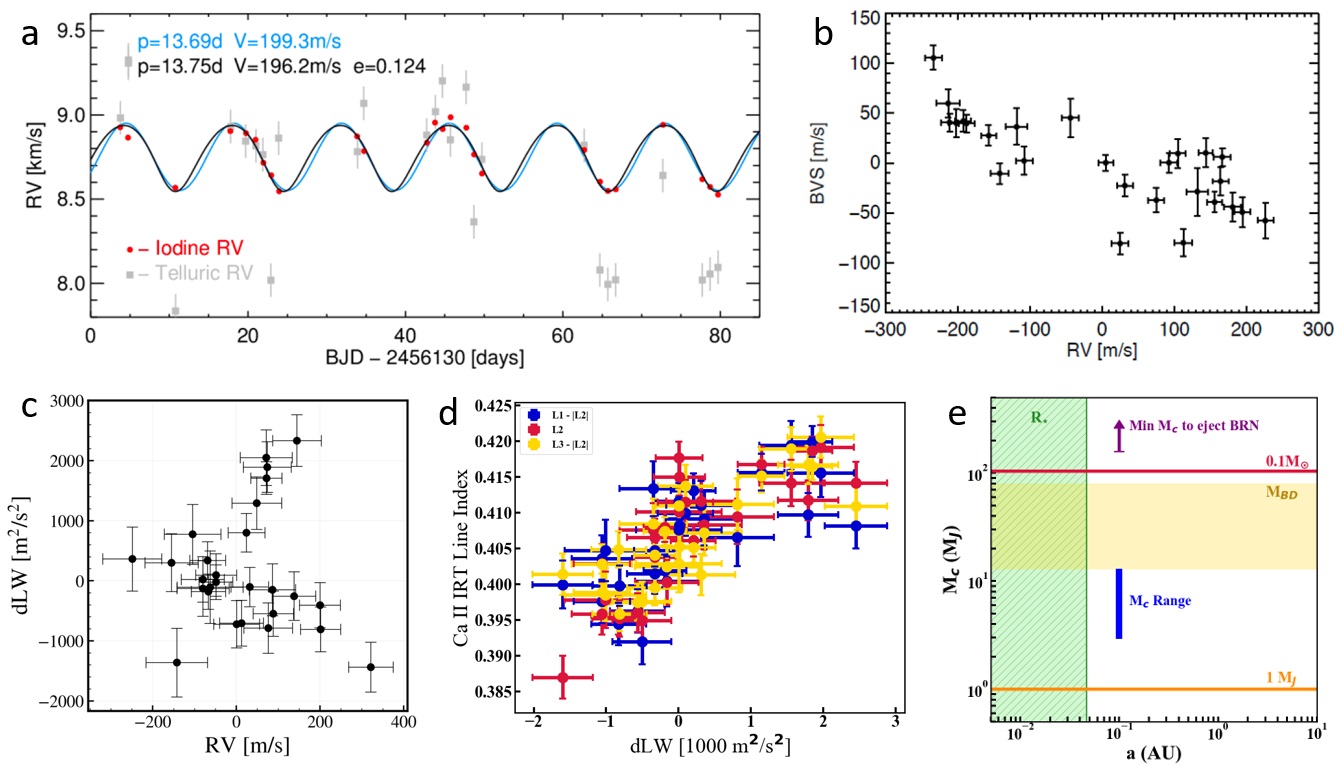}
{\linespread{1.0}
\caption{{\bf The radial velocity of TYC 2597-735-1.} All uncertainties are taken as the standard deviation in each data point. a) The best-fit Keck/HIRES period using the iodine cell calibration technique. Telluric calibration points are presented to show the discrepancy in the two methods. Keck/HIRES RV finds a period of about 13.75 days for a companion that produces an RV amplitude of 196 m/s. b) The bisector velocity span (BVS) as a function of Keck/HIRES RV signal shows an anti-correlation trend. c) HET/HPF differential line width (dLW) as a function of RV, highlighting clear variations in the differential line width as a function of radial velocities, which are observed to vary from -250 m/s to 250 m/s in the HPF RV data. d) Ca II IRT Indices from HET/HPF. L1, L2, and L3 refer to individual Ca II IRT line indices, where L1 = 8500{\AA}, L2 = 8545{\AA}, and L3 = 8665{\AA}. All line indices are normalized to the average index of L2 to display in one figure. dLW versus each line index shows a significant correlation. e) If the Keck/HIRES iodine cell RV signal is the result of a companion, we show the range of mass this companion could have, based on its 13.7-day orbital period (semi-major axis $a \sim$ 0.1 AU; blue vertical line). We also show the minimum mass companion required to eject a collimated, biconical outflow with the velocity of the BRN (purple lower limit), due to the conversion of gravitational energy to kinetic energy as its orbit decays from infinity to $a \sim$ 0.1 AU. We put TYC 2597-735-1’s hypothetical companion into context with other mass objects, including Jupiter (orange line), brown dwarfs (yellow shaded region), and M-stars (0.1 M$_{\odot}$, red line). The current radial extent of TYC 2597-735-1 ($\sim$10 R$_{\odot}$) is shown (green shaded region).}
\label{fig:rv}}
\end{figure}

\begin{figure}
\centering
\includegraphics[angle=0, width=\textwidth]{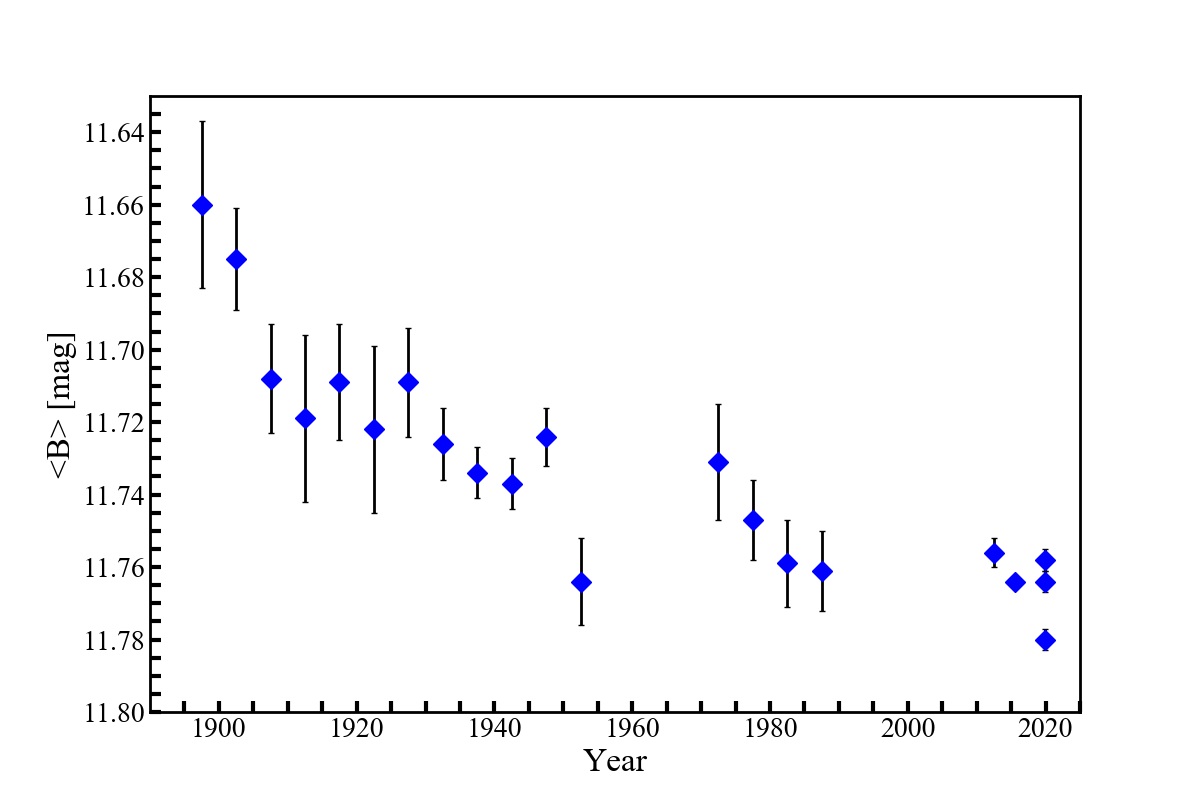} 
{\linespread{1.0}
\caption{{\bf \tyc~light curve since 1895.} A full description of the process of generating \tyc's century-long light curve is provided in the Methods. The uncertainty in the binned magnitudes is the RMS scatter divided by the square root of the number of plates used per bin. The rough trend of \tyc's light curve shows a total B-mag decay of 0.11-0.12 mag between 1895-2015, consistent with 0.09 - 0.1 mag/century. This falls in the range of predicted secular decay in the MESA models for the case study of \tyc's stellar merger history.}
\label{fig:lc}}
\end{figure}

\begin{figure}
\centering
\includegraphics[angle=0, width=0.6\textwidth]{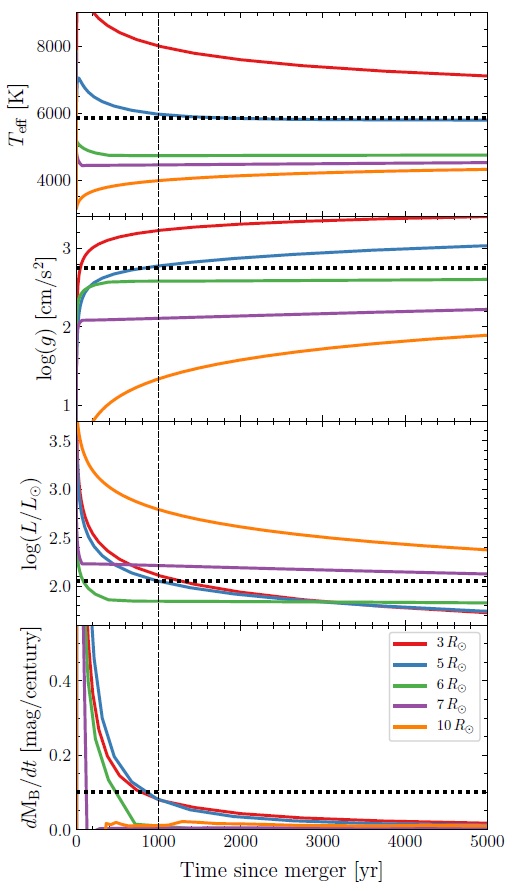}
{\linespread{1.0}
\caption{{\bf The evolution of a stellar merger between a 2M$_{\odot}$ primary star and a 0.1M$_{\odot}$ companion.} MESA evolutionary models were created to look at how the energy injected into the primary star changes its observed characteristics over time. The colored tracks represent mergers at different evolutionary stages of the primary as it evolves towards the red giant branch (RGB). The dashed horizontal lines represent \tyc's observed parameters. This model outcome represents one scenario that helps justify \tyc's history of a stellar merger creating a Blue Ring Nebula 1,000 years later (vertical dashed line).}
\label{fig:mesa}}
\end{figure}

\begin{figure}
\centering
\includegraphics[angle=0, width=\textwidth]{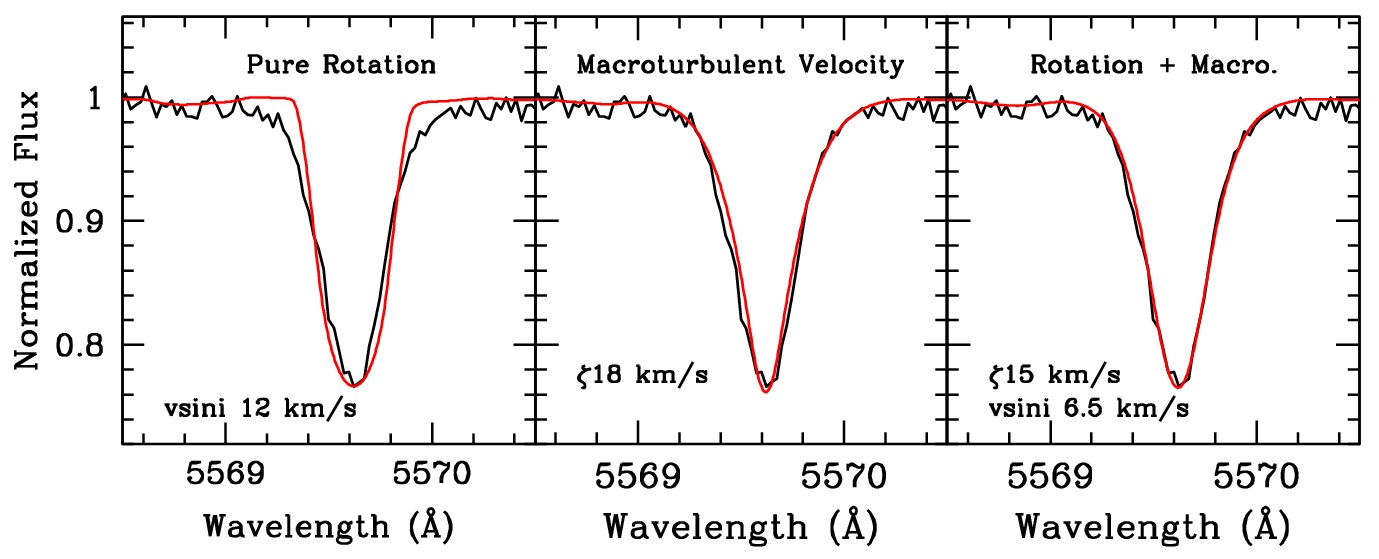}
{\linespread{1.0}
\caption{{\bf Demonstration of the velocity line profile fitting to an unblended Fe~I line (5569.6{\AA}).} (Left) Rotational velocity fit only (red line). The U-shaped rotational velocity profile, alone, does not capture the line wings of the \tyc~Fe~I line line (black). (Middle) Macroturbulence velocity fit only (red line). While the fit to the line wings is improved, the line core is too narrow. (Right) Convolved rotational plus macroturbulent velocity profile provide a better fit to the observed Fe~I line (fit: red line, data: black line). }
\label{fig:rot_velocity}}
\end{figure}

\clearpage
\newpage



\subsection{Supplementary Information}

\subsection{\emph{GALEX} Data Reduction} 
Far-ultraviolet (FUV) and near-ultraviolet (NUV) observations of the ultraviolet nebula (hereby the Blue Ring Nebula or BRN) were taken during the \emph{Galaxy Evolution Explore} (\emph{GALEX}) Deep Imaging Survey (DIS). Each imaging band (angular resolution) covers 1344 –- 1786 {\AA} (4.5$^{\prime\prime}$) and 1771 –- 2831 {\AA} (6.0$^{\prime\prime}$), respectively, with a spatial resolution of 1.5$^{\prime\prime}$ per pixel\cite{Morrissey+05}. The \emph{GALEX} images were taken between 1 - 31 July 2004. All DIS images were corrected using the relative response images of each exposure, then co-added to create the final filter image. FUV images have a total exposure time of 50,287 seconds, while NUV images were taken over a total exposure time of 68,518 seconds. The final FUV and NUV images have sensitivity limits of $\sim$25 m$_{AB}$. A diffuse nebula is observed in the FUV image (m$_{\rm FUV,AB}$ $\sim$ 20.2), with the star TYC 2597-735-1\cite{Hog+00} at the center of the nebula. 

In both \emph{GALEX} FUV and NUV images, starlight from TYC 2597-735-1 and field stars has been removed using a customized source removal routine in the python \texttt{photutils} module. 
The routine masks off the point-spread function (PSF) of any star that falls above a signal-to-noise (SN) threshold (for FUV, SN$>$5, and for NUV, SN$>$2) and replaces the star with a tile of random background noise representative of the surrounding sky. The sky noise is derived from sky background maps provided with each DIS exposure and co-added to create a master sky map. This exercise was done to remove excess flux from the nebular area, to derive the flux from the diffuse nebula only.  

The FUV nebula appears as a ring with a slightly-elliptical shape and a clearly-defined hole at the center. The FUV emission ring of the BRN has an inner spatial extent on the sky d$_{in}$ = 3$^{\prime}$.7 and an outer extent of d$_{out}$ = 7$^{\prime}$.3. The geometry of the BRN is such that the FUV emission shines through where forwards and backwards outflow cones overlap (as drawn in Figure~1(f)). The viewing angle of the nebula was derived by modeling a biconical outflow from a point source and matching the geometry of the overlapping cones to reproduce the profile of the BRN. This gives $i \sim$15$^{\circ}$ (Figure~1(f)). 

\subsection{Multi-Wavelength Look of the UV Nebula} 
Extended Data (ED) Figure~\ref{fig:brn_compilation} shows a compilation of UV, visible light, and infrared images from archival surveys (\emph{GALEX}, DSS-II, 2MASS, and \emph{WISE}) and narrow-band H$\alpha$ (Palomar-COSMIC) of the region around \tyc. DSS-II, 2MASS, \emph{WISE} images were obtained from the IRSA database. Only in the \emph{GALEX} FUV image is the inner thick band of the Blue Ring Nebula visible; no signs of the inner FUV nebula is seen in other visible light or infrared images.  The outline of a shock front can is seen in the \emph{GALEX} FUV, NUV, and H$\alpha$ images. The H$\alpha$ image reveals dual shocks that form two ``loop''-like structures.  The western-most shock in the H$\alpha$ image is associated with one shock loop, while the next filamentary structure moving east (also seen as the western shock in FUV and NUV images) traces the second loop.

\subsection{Properties of the Ultraviolet Nebula} 
The count rate observed by \emph{GALEX} is converted to flux in standard cgs units using the \emph{GALEX} conversion factors for FUV and NUV channels\cite{Hamden+2013}. We find a total flux emitted from the nebula F$_{FUV}$ = 1.5$\times$10$^{-14}$ erg cm$^{-2}$ s$^{-1}$ {\AA}$^{-1}$, corresponding to a total FUV luminosity L$_{FUV}$ = 3$\times$10$^{33}$ erg s$^{-1}$ for a distance of 1.9 kpc to the source (see \emph{Gaia} description below). There is no detectable NUV emission counterpart associated with the nebula, except along the western boundary of the nebula, where H$\alpha$ emission is also detected. We place an upper bound of NUV emission in the nebula of F$_{NUV} <$ 10$^{-18}$ erg cm$^{-2}$ s$^{-1}$ {\AA}$^{-1}$, corresponding to L$_{NUV} <$ 5$\times$10$^{29}$ erg s$^{-1}$. The NUV/FUV luminosity ratio of the BRN is $<$1.5$\times$10$^{-4}$; there are $>$10$^4$ more FUV photons emitting from the smooth, uniform ring of the BRN than NUV photons.

Along the western (left-hand side) edge of the FUV nebula, a shock filament is observed in emission in both the FUV and NUV (Figure~1). Since we do not capture all of the shock emission in FUV or NUV emission, just one part of it, we present the average flux in both bandpasses, rather than the total. The average flux measured in this shock filament is F$_{FUV}$ $\sim$ 10$^{-18}$ erg cm$^{-2}$ s$^{-1}$ {\AA}$^{-1}$ and F$_{NUV}$ $\sim$ 6$\times$10$^{-19}$ erg cm$^{-2}$ s$^{-1}$ {\AA}$^{-1}$; this shows less than an order of magnitude difference in FUV and NUV flux along the shock, versus over four orders of magnitude difference in flux within the ring of the BRN itself. The average shock luminosity in the FUV and NUV are L$_{FUV} \approx$ 5.2$\times$10$^{28}$ erg s$^{-1}$ and L$_{NUV} \approx$ 1.3$\times$10$^{29}$ erg s$^{-1}$, resulting in a FUV/NUV luminosity ratio along the shock $\sim$0.4. Less than unity, this FUV/NUV ratio is more consistent with the expected \emph{GALEX} FUV/NUV ratio for two-photon emission (0.37)\cite{Bracco+2020} than dust scattering ($>$1)\cite{Murthy+2010}.

\subsection{Emission source of the Blue Ring Nebula} 
\emph{GALEX} FUV grism spectra of the BRN were obtained between 23 July - 3 August 2007, corresponding to a total observing time of 14,950 seconds. The FUV grism image reveals that the BRN shines in the wavelength range $\sim$1550 - 1650{\AA}, which is consistent with the emission processing being collisionally-induced molecular hydrogen fluorescence (ED Figure~\ref{fig:fuv_grism}). 
We note that an exact bandpass constraint is impossible with the slitless FUV grism spectrum, given the scale of the BRN, which spreads the emission out and smears the inferred spectrum across the spatial extent of the emission region. Molecular hydrogen fluorescence (H$_2$F) is a natural explanation for the BRN, as the fluorescence is confined to FUV wavelengths only - it has no emission counterpart in the NUV ($\lambda$(H$_2$F) $<$ 1700{\AA}). If an appreciable amount of any other atomic/ionic species resides within the nebula, the presence of a NUV emission counterpart would be expected. For example, some carbon species have strong FUV emission coincidence with H$_2$ fluorescence (e.g, CIV 1548,1550), but would also appear with an emission counterpart in the NUV (e.g., CIII] 1906, CII] 2324).

\subsection{Neutral Hydrogen Mass in the BRN} 
The FUV luminosity, assuming the emission is exclusively produced by an optically-thin layer of H$_2$, can be converted to a rate of H$_2$ fluorescence over the entire BRN. This conversion assumes that the nebular gas is very thin, such that the molecular density does not saturate the absorption cross-section for excitation. 

To justify this assumption, we estimate the volumetric density of H$_2$ ($n$(H$_2$)) in the BRN. To do so, we estimate the rate that H$_2$ is collisionally excited by hot electrons in the medium, assuming that the source of H$_2$ excitation is associated with a reverse shock through the BRN (as is shown in the last panel of Figure~3). The nebula itself (excluding the shock filament) emits a total H$_2$ luminosity L$_{H_2}$ = 3$\times$10$^{33}$ erg s$^{-1}$. We convert this to a total rate of H$_2$ collisions required to produce this luminosity, $\langle \Gamma_{H_2 + e^-} \rangle$ = L$_{H_2}$/E(H$_2$F) $\sim$ 2.5$\times$10$^{44}$ collisions s$^{-1}$, where $\Gamma_{H_2 + e^-}$ is the rate of H$_2$ fluorescence by electron collisions and E(H$_2$F) is the energy of the transition to produce H$_2$ fluorescence. The emitting surface area of the BRN is quite large (A$_{BRN}$ $\sim$ 10$^{39}$ cm$^2$), so we calculate the average expected collision rate with electrons over the area of the nebula using derived cross sections of collisions between H$_2$ and free electrons: $\langle \gamma_{H_2 + e^-} \rangle$ = $\langle \Gamma_{H_2 + e^-} \rangle$ / ($A_{BRN}$ / $\sigma$(H$_2$)) $\sim$ 10$^{-11}$ collisions/sec, where $\sigma$(H$_2$) $\sim$ 10$^{-16}$ is the cross-section between H$_2$ and warm-hot electrons (E$_{e^-}$ $>$ 20 eV\cite{Liu+96}). The expected density rate of collisions between electrons and cool-warm (T $\sim$ 500 K) H$_2$ is $C_{H_2 + e^-}$ $\sim$ 10$^{-11}$ cm$^3$ s$^{-1}$\cite{Prasad+Huntress+80b}. Setting the rates of collisions between warm H$_2$ and free electrons and the observed rate of H$_2$ fluorescence, we find: $\langle \gamma_{H_2 + e^-} \rangle$ = $C_{H_2 + e^-}$ $\times$ $n$(H$_2$), which means $n$(H$_2$) $\sim$ 1 cm$^{-3}$. This density is consistent with the diffuse, electron-impact H$_2$ fluorescence observed in Mira's FUV tail\cite{Martin+07}.

With the assumption that the H$_2$ nebula is quite diffuse, we estimate how much mass of \emph{neutral} hydrogen has been produced in the visible area of FUV emission exposed by this H$_2$ fluorescence process; each time an H$_2$ molecule is excited collisionally by an electron, it has a non-trivial chance (10-15\%) to dissociate into neutral hydrogen (HI)\cite{Abgrall+1997}. Assuming that the total number of H$_2$ fluorescence transitions equals the total number of molecules undergoing these transitions, the total amount of H$_2$ emitting per second in the BRN $\sim$ 2.5$\times$10$^{44}$ molecules s$^{-1}$. We convolve the rate of transitions with model H$_2$ fluorescence spectra, assuming collisional excitation via electrons, and use knowledge of the transition strengths and molecular dissociation probabilities per upper electronic level cascade to ground electron level\cite{Abgrall+93,Abgrall+1997} to estimate the rate of H$_2$ dissociation from the BRN FUV luminosity. We find the rate of H$_2$ dissociation $\gamma_{H_2 - HI}$ $\sim$ 2.5$\times$10$^{-14}$ solar masses (M$_{\odot}$) per second. 
If the H$_2$ dissociation rate in the BRN has been constant over the age of the BRN (see Age discussion below), then we can estimate the mass of HI in the BRN as: M$_{H_2 - HI}$ $\sim$ $ \gamma_{H_2 - HI} \times$ t$_{BRN}$. This places an upper limit on the total mass of the nebula that has been dissociated from H$_2$ to HI, M$_{H_2 - HI} \lesssim$ 0.0008 - 0.004 M$_{\odot}$ = 0.8 - 4.1 Jupiter masses (assuming t$_{BRN}$ between 1,000 - 5,000 years; see Age estimate below).

\subsection{Palomar/COSMIC H$\alpha$ Shock Rings} 
After the initial discovery of the Blue Ring Nebula with \emph{GALEX}, follow-up observations of the nebula were taken in H$\alpha$ to search for nebular emission from neutral hydrogen. H$\alpha$ narrow-band images were taken on 10 September 2004 with the Palomar Hale 200-inch telescope using the COSMIC instrument. A total exposure time of 2,700 seconds (3 exposures of 900 seconds each) was taken, and R-band images for continuum subtraction were taken alongside H$\alpha$ narrow-band images (300 sec per exposure in R-band). Only the associated shock front of the Blue Ring Nebula is detected in H$\alpha$ (Figure~1, ED Figure~\ref{fig:brn_compilation}).

\subsection{Keck/LRIS Observations of Shock Filaments} 
The optical filaments bright in H$\alpha$ were observed on the nights of 23 and 24 July, 2006 with the Low Resolution Imaging Spectrograph (LRIS) on the Keck I telescope using a slit-mask with nine apertures of 0.7 arcsecond width on the brightest knots in the filaments.  Keck/LRIS was configured with the 600/4000 grism in the blue and with the 1200/7500 grating in the red and no filters on either channel, but using the 500 dichroic.  The grating angle was tuned to optimize throughput for the H$\alpha$ line on the red side, while H$\beta$ was the primary line on the blue side.  The resolution element resulting from these configurations is roughly 2 - 3 {\AA} full width half maximum (FWHM), giving a velocity resolution of about 100 km/s.  Four 2,400-second exposures were obtained on each of the two nights under good conditions and at low ($<$1.1) airmass. Each night was reduced and calibrated separately and combined to create a master spectrum for each of the knots.  The wavelengths were calibrated using internal Neon calibration lamps between each exposure and also adjusted using night sky lines and corrected to the heliocentric system. Once corrected, the H-Balmer series line spectra in each LRIS slit were converted from wavelength- to velocity-space. 

Selective slit spectra along the western boundary of the nebula observe hydrogen-Balmer series emission along the shock boundary. Figure~1(d,f) shows the placement of the LRIS slits in both the H$\alpha$ narrow-band image and geometry schematic of the BRN and the resulting H$\alpha$ velocity shifts from line center (in velocity space) as observed over different parts of the two shock loops. Maximal offsets reach $\pm$400 km/s, which we interpret as the highest velocity of the shock filament associated with the BRN.  We measure a mean H$\alpha$ flux along the western filament F(H$\alpha$) = 9.7$\pm$3.0 $\times$ 10$^{-17}$ erg cm$^{-2}$ s$^{-1}$ {\AA}$^{-1}$ and a mean H$\beta$ flux through the same LRIS slits along the western shock of 1.5$\pm$0.4 $\times$ 10$^{-16}$ erg cm$^{-2}$ s$^{-1}$ {\AA}$^{-1}$. We measure an average H$\alpha$/H$\beta$ = 0.74$\pm$0.29 along this shock, indicating a low H$\alpha$/H$\beta$ ratio inconsistent with local thermodynamic equilibrium. Instead, the ratio is more consistent with those observed in solar flares (H$\alpha$/H$\beta$ $\sim$ 1), which are interpreted as being non-LTE and significantly impacted by charge exchange with free electrons\cite{Zirin+1982,Capparelli+2017}. The optical and UV emission observed along the shock are therefore consistent with a non-radiative shock front with recombination of hydrogen dominating the emission from this region\cite{McKee+1980}.

\subsection{Constraining the Age of the Blue Ring Nebula} 
The maximum velocity of the BRN forward shock places constraints on the age of the nebula. Assuming that, at early times after the merger, the ejecta coasts at a constant velocity matching that of the forward shock, $v_{\rm sh}$, the source age can be estimated as $t = r_{\rm sh}/v_{\rm sh}$.  As the ejecta begins to sweep up an appreciable mass of the external interstellar medium (ISM) comparable to its own, a reverse shock will pass through the ejecta.  At late times, the evolution of the forward shock will approach that of Sedov-Taylor blast wave, for which $r_{\rm sh} \propto t^{2/5}$ in the case of a constant density ISM and, therefore, $v_{\rm sh} = dr_{\rm sh}/dt = (2/5)r_{\rm sh}/t$, implying a source age $t = (2/5)r_{\rm sh}/v_{\rm sh}$.  The age of the source can therefore be constrained between
\be
\frac{2}{5}\frac{r_{\rm sh}}{v_{\rm sh}} < t < \frac{r_{\rm sh}}{v_{\rm sh}},
\ee
which comes out to be 2,000 years $<$ $t$ $<$ 5,000 years, given the inferred size and expansion rate of the BRN. As demonstrated in our MESA models (see below), the closest-matched model to the current properties of the BRN central start, \tyc, is best suited to an age $t \sim$ 1,000 years since the stellar merger began. Therefore, we adopt an age range for the BRN between 1,000 years $<$ $t$ $<$ 5,000 years throughout the course of this study.

\subsection{{\it Gaia} distance} 
The {\it Gaia} DR2 parallax of \tyc~is 0.518$\pm$0.029 mas, from which employing a Bayesian method with the preferred ``exponentially decreasing space density'' (EDSD) prior and a distance scale corresponding to a disk population of 1000 pc at a galactic latitude\cite{Bailer-Jones+15} of +39.4$^{\circ}$, we obtain a distance of 1935$^{+127}_{-91}$ pc. 
The fractional error on the parallax is sufficiently small that the details of the method and the adopted prior only change the distance by less than one-tenth of the quoted 1$\sigma$ error bars on the distance.

\subsection{Archival Photometry \& Spectral Energy Distribution of \tyc} 
We compile the spectral energy distribution (SED) of \tyc~using pre-existing photometric observations taken on a variety of UV, optical, and infrared platforms (Figure~2).  The ultraviolet portion of the spectrum is taken from the \emph{GALEX} FUV and NUV images.  The optical portion of the SED comes from the Pan-STARRS\cite{Chambers+16} and Sloan Digital Sky Survey\cite{Gunn+06} surveys.  Archival 2MASS, AKARI, and WISE data provide the near- to mid-IR coverage.  We find no clear detection of \tyc~in IRAS far-IR archival data, providing only upper limits in the far-infrared.

\textbf{Interstellar and Circumstellar Reddening} 
The extinction maps and IRSA far-IR photometry towards \tyc~indicate a mean E(B-V) of 0.0181 magnitudes in a 5 arc-minute box centered on \tyc\cite{Schlegel+98}, with a minimum E(B-V) of 0.0167 mag and a maximum E(B-V) of 0.0204 mag.  The reddening maps are based on 100 micron dust emission and represent the maximum interstellar reddening in a given direction\cite{Schlegel+98}; they do not provide information on circumstellar reddening. 

A second estimate of the photometric reddening towards \tyc~was made, including any circumstellar reddening, by analyzing the interstellar Na~I~D lines.  The equivalent width (EW) of foreground Na D1 and Na D2 absorption has been shown to be linearly proportional to extinction\cite{Poznanski+12}.  We estimate the maximum EW of the D1 and D2 features to have EW(D1) $\sim$ 68 m\AA and EW(D2) $\sim$ 110 m\AA. These EWs indicate E(B-V)=0.023 mag for toward TYC 2597-735-1\cite{Poznanski+12}, consistent with the IRSA reddening maps\cite{Schlegel+98} upper limit. 
We adopt E(B-V)=0.02 mag and estimate its photometric properties from the corrected color-temperature relations.  

\textbf{Spectral Energy Distribution} 
\tyc~exhibits photospheric continuum emission from the UV to J-band consistent with a stellar blackbody emission with an effective temperature T$_{eff} \sim $5,850 K.  We fit ATLAS9 synthetic stellar models\cite{CastelliKurucz2003} to the spectral energy distribution (SED) of \tyc, using stellar parameters close to those derived for \tyc~(see {\bf Stellar Properties} for more details). The ATLAS9 catalog provides a pre-set grid of synthetic stellar spectra with a defined range of stellar parameters. We fit stellar models for T$_{eff}$ = 5,750K and 600K (the grids go through T$_{eff}$ space in steps of 250K), log(g) = 2.5 and 3 (the grids go through log(g) space in steps of 0.5), and [M/H] = -1.0.

We find that, from the NUV to NIR, the synthetic stellar models fit well to the observed photometry, through we note a stark contrast between the observe FUV flux and that of the stellar model. For a star with the properties close to those reported for \tyc, we find that stellar synthesis models predict a range of FUV flux output by the star $F_{\rm 1500} \sim$ 10$^{-8}$ - 10$^{-7}$ Jy (2.6 $\times$ 10$^{-19}$ - 10$^{-18}$  erg cm$^{-2}$ s$^{-1}$ {\AA}$^{-1}$, or $\sim$29 - 26.5 AB mag). However, \emph{GALEX} FUV imaging observes \tyc~to have a FUV flux $F_{\rm 1500, obs} \sim$ 2.0 $\times$ 10$^{-5}$ Jy (2.6 $\times$ 10$^{-16}$ erg cm$^{-2}$ s$^{-1}$ {\AA}$^{-1}$, or 20.5 AB mag); we observe $> 10^{2} \times$ more flux in the FUV than the stellar models predict ($\sim$6-8 AB mag).

Excess FUV emission (both in emission lines and continuum) are well-known and reported in systems with accretion disks\cite{Ardila+2013, France+2014, Sahai+2015}. \emph{GALEX}-FUV imaging covers a wavelength range (1350 - 1750{\AA}) where studies report excess FUV emission produced in accreting systems accounts for a small percentage ($<$5\%) of the total measured FUV excess\cite{France+2014}. Using the total accretion luminosity derived from the observed H$\alpha$ emission (see {\bf Stellar H$\alpha$ Emission}), the measured FUV excess accounts for 3\% of the total accretion luminosity. This is well within reason, given that the majority of \tyc's FUV excess is likely not measured (i.e., shorter wavelengths than were observed with \emph{GALEX}).

From 2$\mu$m out to longer wavelength, the SED exhibits excess infrared emission relative to the single temperature stellar continuum.  This IR excess likely arises from a dusty, warm circumstellar disk, similar to those seen around protostars\cite{Hartmann+98} and some post-AGB systems\cite{vanWinckel03}.  A disk-like geometry for the dust, which lies in a plane perpendicular to our line of sight (and hence approximately perpendicular to the symmetric axis of the nebula), is further supported by the lack of circumstellar reddening of \tyc, inferred from our Na I D analysis (see {\bf Interstellar and Circumstellar Reddening}).  The IR excess peaks at $\lambda \sim$8 $\mu$m, corresponding to an average dust characteristic temperature $\sim$600 K. 

If this temperature reflects re-radiated light from thermal dust particles in orbit around \tyc, then the dust peak location is $\sim$2.2 AU (depending on the albedo of the dust, which is entirely dependent on the dust grain distribution, which ranges between 0.2 - 3 AU\cite{Mattila1979,Jura+1998,Mulders+13}). The dust disk area is derived by modeling the dust distribution over a thin layer in the circumstellar disk to reproduce the IR SED\cite{Jura03}. We verify that r$_{in} \sim$0.25 AU and r$_{out} \sim$2.8 AU best reproduce the IR excess SED (Figure~2(a)). We empirically derive the dust dist mass from the infrared spectral energy distribution of \tyc, assuming all the dust is roughly the same temperature in a thin layer at the equilibrium distance and that the mean dust size is small\cite{Jura+1998,Chen2005}. We assume an average dust temperature T$_{dust}$ = 600 K and take the equilibrium distance $D_{\rm dust} \sim$2.2 AU. The surface density of the dust disk is taken to be small ($\rho_{\rm dust} \approx$2.5 gm/cm$^3$\cite{Chen2005}) with an average small dust particle size ($a_{\rm dust} \sim$10$^{-4}$ cm, or $\sim$1 $\mu$m) to ensure we are estimating a lower limit to the dust disk mass.  Using the following relation\cite{Jura+1995},
\be
M_{\rm dust} \approx \frac{16 \pi}{3} \times \frac{F_{\rm IR}}{F_{\star}} \times \rho_{\rm dust} \times D_{\rm dust}^2 \times a_{\rm dust}
\ee
we derive a minimum dust mass in the present-day circumstellar disk $M_{\rm dust} \gtrsim 5 \times 10^{-9} M_{\odot}$.

\subsection{Keck/HIRES Observations} 
High-resolution stellar spectra of \tyc~were obtained at optical wavelengths using the Keck High Resolution Echelle Spectrometer (HIRES) with the C2 Decker and iodine cell calibration\cite{Howard+10}. The purpose of these spectra were to search for and measure a radial velocity signal from \tyc~to determine if its unusual characteristics were attributed to a close-in companion, but the Keck/HIRES spectra were also paramount in investigating the properties of \tyc~itself (see {\bf Stellar Properties}). TYC 2957-735-1 was observed 26 times over 75 days between 25 July - 9 October, 2012. The typical observation achieved a S/N of 30 per pixel (60 per resolution element). An iodine free template spectra of 900 seconds achieved a spectral resolution of R=171,000 and S/N of $\sim$180 per pixel at 5500{\AA}. The median error of the iodine calibrated velocities is 11.3 m/s.

\subsection{Radial Velocity Methods \& Results}
The systemic velocity of TYC 2957-735-1 is 8.74 km/s\cite{Gaia+18}. The iodine calibrated measurements show a detectable periodic radial velocity of semi-amplitude 199 m/s and period of 13.7 days assuming a circular orbit. Allowing eccentricity as a free parameter yields a semi-amplitude of 196 m/s, a period of 13.8 days and eccentricity of 0.124 (ED Figure~\ref{fig:rv}(a)). 

Hypothetically, if the measured Keck/HIRES RV using the iodine cell cross-correlation is linked to a close-in companion with $\sim$13.7 day orbital period, we argue that this companion is not sufficiently massive enough to eject the BRN we observe today. A companion with P = 13.7 days would have a semi-major axis $a \sim$0.1 AU. Assuming \tyc~has a mass 2 M$_{\odot}$ and the orbital plane ranges from perpendicular to the axis of the Blue Ring Nebula ($i_{\rm min} \sim$ 15$^{\circ}$) to edge-on to our vantage point ($i_{\rm max} \sim$ 90$^{\circ}$), the companion mass will be between $M_{\rm p} \approx 3.1-12.1 M_{\rm J}$ (ED Figure~\ref{fig:rv}). The total gravitational energy liberated as the planet inspiraled from $\infty$ to its final semi-major axis $a_{\rm p}$ is only 
\be
\Delta E \simeq \frac{GM_{\star}}{a_{\rm p}} \approx 10^{45}{\rm erg}\left(\frac{M_{\rm p}}{10M_{\rm J}}\right)\left(\frac{a_{\rm p}}{0.1{\rm AU}}\right)^{-1} \approx 10^{45}{\rm erg},
\ee
which is at least an order of magnitude smaller than the minimum kinetic energy needed to release the BRN ($E_{\rm KE} = M_{\rm ej}v_{\rm ej}^{2}/2 \gtrsim 10^{46}{\rm erg}$, assuming $M_{\rm ej}$=0.01M$_{\odot}$). 
Thus, even if the RV periodicity were from an orbiting companion which underwent an earlier phase of common envelope evolution with the primary star, the maximum energy released in bringing the companion to its present orbit would not be sufficient to explain the observed properties of the BRN. We assume the inclination of the hypothetical RV orbit is $i$ = 15$^{\circ}$ (where $i$ = 90$^{\circ}$ is edge-on and $i$ = 0$^{\circ}$ is a face-on orbit), corresponding to our cone-axis inclination in our biconical model of the nebula. Only in a specific geometry where the RV orbit is completely face-on (0$^{\circ} < i < $2$^{\circ}$) does the companion mass begin to approach where the energy released via gravitational insprial could release an outflow with enough kinetic energy to explain the BRN ($\sim$ 0.1 $M_{\odot}$).  

We performed a bisector velocity span (BVS) analysis on the Keck/HIRES iodine cell-calibrated RV results, which is meant to provide a test as to whether RV signals measured from a star are tied to stellar activity, like star spots or atmospheric pulsations\cite{Boisse+2011,Figueira+2013}. We find a statistically significant anti-correlation between the BVS and iodine radial velocities (Spearman's rank correlation coefficient = -0.76, p-value = 7.5$\times$10$^{-6}$; ED Figure~\ref{fig:rv}), which favors the RV signal being produced by stellar activity.  H$\alpha$ emission and variability (see {\bf Stellar H$\alpha$ Emission}), measurable macroturbulence affecting \tyc's stellar line profiles (see {\bf Rotational Velocity \& Macroturbulence}), and the presence of a dusty  circumstellar disk around \tyc~(see {\bf Spectral Energy Distribution}) all provide additional support that stellar activity is the main culprit of the observed HIRES iodine cell RV signal. If \tyc~is accreting material from a disk as suspected from these diagnostics, then it is reasonable to conjecture that the atmosphere is disturbed or that hot spots near the poles exist\cite{Figueira+2013}. Indeed, the period of the HIRES RV is not consistent with that of the rotational velocity of \tyc~(6.5 km/s, corresponding to P$\sim$20 days), so linking the activity near the stellar poles, where the period of revolution may better match the 13.8 days, seems a plausible explanation. 

Moreover, we find that radial velocities from the same HIRES data measured with the telluric line method of Chubak et al. 2012\cite{Chubak+2012} do not agree with the iodine cell calibrated measurements, as shown in ED Figure~\ref{fig:rv}. There are differences of up to 600 m/s, whereas the telluric method claims to be accurate to within $\sim$100 m/s. We do not attempt a periodic fit to the telluric derived velocities. The reason behind this significant disagreement is not known. and it raises concerns regarding the reliability of either calibration method. 

To provide an ancillary interpretation to the Keck/HIRES RV results, we obtained additional RV data in the near-infrared with the Habitable-zone Planet Finder (HPF) to determine if the iodine based periodicity persists at longer wavelengths.

\subsection{HET/HPF Observations and RV reduction} 
HPF is a high resolution ($R\sim55,000$) near-infrared spectrograph\cite{mahadevan2012,Mahadevan+2014} that is actively temperature stabilized to the milli-Kelvin level\cite{stefansson2016}, and covers the $z$, $Y$, and $J$ bands spanning 810-1280 nm. Between 26 April -- 23 July, 2018, we acquired 27 spectra of \tyc~using HPF. 20 of the spectra were obtained with an exposure time of 960s with a mean Signal-to-Noise (S/N) = 127 per 1D extracted pixel at $\lambda = 1000 \unit{nm}$, and 7 of the spectra were obtained with an exposure time of 630 seconds with a mean S/N = 106 per 1D extracted pixel at $\lambda$ = 1000 nm.

HPF has a near-infrared (NIR) laser-frequency comb (LFC) calibrator which has been shown to enable $\sim$20 cm/s calibration precision in 10-minute bins, and 1.5 m/s RV precision on-sky on the bright and stable M-dwarf Barnard's Star over months\citep{metcalf2019}. Due to the faintness of the target, we elected to not use the simultaneous LFC reference calibrator for these observations to minimize any possibility of scattered LFC light in the target spectrum\cite{stefansson2020}. Instead, the drift correction was performed by extrapolating the wavelength solution from other LFC exposures on the nights of the observations\cite{stefansson2020}. This methodology has been shown to enable precise wavelength calibration and drift correction down to the $\sim$30 cm/s level, much smaller than the RV variability observed in this system.

The HPF 1D spectra were reduced and extracted with the custom HPF data-extraction pipeline\cite{ninan2018,kaplan2018,metcalf2019}. After the 1D spectral extraction, we extracted precise NIR RVs\cite{stefansson2020}, using seven of HPF's cleanest (of tellurics) orders, spanning a wavelength range between 854-900 nm and 994-1058 nm. To extract the RVs, we use a modified version of the SpEctrum Radial Velocity Analyzer (SERVAL) code\cite{zechmeister2018}, which uses the template-matching technique to extract precise RVs, adapted for use for the HPF spectra. The stellar activity indicators used in this work -- including the differential line width indicator (dLW), chromatic index (CRX), and Calcium Infrared Triplet (Ca IRT) indices -- are the same as defined in the SERVAL pipeline\cite{zechmeister2018}.

We also see evidence of activity-induced RV variations in the HET/HPF RVs, which we interpret as features intrinsic to the star which could include stellar surface inhomogeneities (e.g., starspots, plages, and/or other active regions) and/or stellar pulsations in the evolved star. From the HPF RVs, we see clear RV variations with an amplitude of $\sim$200-250 m/s in the 13-visit sampling. ED Figure~\ref{fig:rv}(c) shows evidence of clear variations in the HPF differential line width (dLW) indicator as a function of the RVs—which although not a clear 1:1 correlation—the large variations are suggestive of complex activity effects being the source of the RV variations. To this end, clear variations are further seen in the Ca~II infrared triplet (IRT; 8500{\AA}, 8545{\AA}, 8665{\AA}) indices, which closely correlate with the dLW indicator (ED Figure~\ref{fig:rv}(d)). While the dLW traces variations in the mean stellar line profile, the Ca II IRT triplet traces activity in the chromosphere of the star. As such, the strong correlation seen in ED Figure~\ref{fig:rv}(d) (correlation coefficient $>$ +0.75, p-value $<$ 10$^{-5}$) hints at a possible link between chromospheric and line-profile effects, which could be explained as active regions rotating in and out of view and/or polar stellar hot-spots evolving at a characteristic timescale of 13.75 days.

We further note that with our current sampling with HET/HPF -- which covers approximately half of the phase of the 13.75-day periodicity seen in the Keck/HIRES RVs -- the amplitude of the RV variations in HPF likely represents a lower limit on the NIR RV amplitude. Generally, activity-induced RV variations are expected to show lower amplitudes in the NIR than in the optical\cite{Marchwinski+2015}. We thus speculate that given the 5 year baseline between the HIRES and HPF RVs, that the stellar surface-features could have evolved to give rise to larger activity-induced variations seen during the HPF RV observations, although continued long-term RV observations across the optical and NIR are required to further test this hypothesis. 

With the combined evidence in place from two independent RV surveys, we suspect that the RV signals observed by Keck/HIRES and HET/HPF are dominated by stellar surface activity and modulations on \tyc. Even if a close-in binary companion does exist, we argue that it is not sufficiently massive enough to explain the mass ejection that happened at \tyc~$>$1,000 years ago that is now seen as the BRN.

\subsection{Stellar H$\alpha$ Emission} 
For what appears to be a red sub-giant star (see below), \tyc~displays a highly unusual characteristic: H$\alpha$ line emission in its stellar spectrum. H$\alpha$ emission from star systems can be produced by a variety of physical mechanisms, including accretion (e.g., T Tauri stars, Herbig Ae/Be, symbiotic stars), magnetic activity (e.g., low-mass stars), stellar rotation (e.g., active M-dwarfs), and stellar pulsations (e.g., Mira variables pulsating and heating part of their stellar atmosphere) - all systems with heightened levels of stellar surface activity\cite{Bertout89, Witham+06, Hamilton+12}. In many cases, H$\alpha$ emission is observed to be variable (e.g., shape, flux) over time (although the timescale of variability is not typically well characterized\cite{Hamilton+12}) and may display both emission and absorption features (or periodic emergence of H$\alpha$ emission).  \tyc's H$\alpha$ emission does not appear to coincide with a stellar H$\alpha$ absorption component and has been observed in emission during all observations of \tyc~between 2004 and 2014. The average flux of the H$\alpha$ emission line over this time is F(H$\alpha$) $\sim$ 10$^{-13}$ erg cm$^{-2}$ s$^{-1}$, implying a luminosity L(H$\alpha$) $\sim$ 5$\times$ 10$^{31}$ erg s$^{-1}$.

As shown in Figure~2(b) and ED Figure~\ref{fig:halpha_bisector}, the H$\alpha$ line profile shape shows variability over timescales of days. One H$\alpha$ observation occurred in 2004, with more regular follow-up with Keck/HIRES taken through 2012 as a part of the Keck/HIRES RV campaign.  The H$\alpha$ line profile tends systematically towards bluer wavelengths throughout 2012 (ED Figure~\ref{fig:halpha_bisector}), while the 2004 profile is fairly symmetric and centered on the line center.  In the 2012 line profiles, the peak of the line emission actually fluctuates around H$\alpha$ line center, but the bisector values along different points of the line profile (see below) all tend towards bluer velocity shifts away from line center. This blue-shifting line profile behavior suggests material is flowing onto \tyc\cite{Lima+10}, supporting the idea that \tyc~may be interacting and accreting material from a circumstellar disk. 

We quantify the deviation of the H$\alpha$ line center by calculating the line profile bisector at various points in the emission line profile.  Bisectors were measured at 10\% intervals of line peak. All line profiles were normalized with continuum levels matched, to best compare the line profile behavior at similar positions in the profile shape. At two points in the profiles - where the line profile is most sensitive to the line wings (continuum level - 10\% of the peak flux) and near the half flux point of the profile ($\sim$50\% - 70\% of the peak flux), the deviation from line center bisector velocity shift $<$ -5 km/s, with an average $\sim$ -10 km/s (ED Figure~\ref{fig:halpha_bisector}).  

H$\alpha$ emission in young stars is widely associated with material from a circumstellar disk being funneled onto the star via magnetospheric accretion, or removed from the disk in the form of winds\cite{Lima+10}. A similar reservoir of material could be fueling the H$\alpha$ emission and variability of \tyc. As discussed, \tyc~exhibits an infrared excess that supports the presence of a circumstellar disk orbiting the star, which is an expected consequence in stellar merger simulations\cite{MacLeod+18}. This disk provides the necessary material that would accrete onto \tyc~through magnetospheric accretion, creating signposts of stellar activity. If L(H$\alpha$) is a direct measure of the accretion luminosity, L$_{\rm acc}$\cite{Barentsen+11}, then L$_{\rm acc}$ $\sim$ 10$^{33}$ ergs s$^{-1}$. Converting L$_{\rm acc}$ to mass accretion rate ($\dot{M}$), \tyc~accretes $\lesssim$1.5$\times$10$^{-7}$ M$_{\odot}$/yr of material. We note that converting L(H$\alpha$) to a mass accretion rate provides an \emph{upper limit} for the accretion rate, as studies have shown that H$\alpha$ emission is also generated by disk material itself, disk winds and other sources embedded in disks\cite{Lima+10,Patel+17,Thanathibodee+19}.

\subsection{Stellar Properties of \tyc} 
We present a brief description about the determination of \tyc's stellar properties using Keck/HIRES spectra. 

\textbf{Luminosity and Radius of \tyc} 
The Johnson V-band magnitude, derived from SDSS gr photometry\cite{Jester+2005}, is 11.149 mag. Using extinction laws\cite{Winkler+1996} and E(B$-$V)=0.02 mag, the V-band magnitude corrected for extinction is V$_0$ = 11.087 mag. The absolute V-band magnitude of BRN, based on the \emph{Gaia} parallax, is then M$_{V}$ = -0.34$\pm$0.12 mag. From bolometric corrections\cite{Kurucz1979} for the Sun and \tyc, -0.193 and -0.219 respectively, we derive log(L/L$_{\odot}$) = 2.07 for \tyc\cite{Torres+2010}.  Paired with the adopted T$_{eff}$ = 5850 K and log(L/L$_{\odot}$) = 2.07, we find R = 10.5 $\pm$ 0.5 R$_{\odot}$ for \tyc.

\textbf{Rotational Velocity \& Macroturbulence} 
Based on model atmosphere parameters, derived here for TYC-2597-735-1, and the spectrograph slit function of Keck/HIRES, we performed spectrum synthesis calculations to match line profiles in our HIRES spectrum. We found that the U-shaped profiles characteristic of pure rotational broadening failed to reproduce the observed line profiles, particularly in the red and blue wings of each line.  Conversely, a Gaussian macroturbulent velocity distribution fit the line wings, but not the line cores.  We demonstrate what these line fits look like in ED Figure~\ref{fig:rot_velocity}.  
The final solution constrained by line full width half max (FWHM), as well as the line cores and wings, was obtained for a $v sin i$= 6.5 km/s combined with a macroturbulent velocity parameter of 15 km/s. Assuming the rotation axis of \tyc~aligns with the symmetry axis of the BRN ($i \sim$15$^{\circ}$), we derive a de-projected surface rotational velocity $v_r \approx 25$ km/s. 

We find this de-projected surface rotation velocity to be larger than expected for a star which has just evolved off the main sequence ($v <$ 10 km/s)\cite{deMedeiros+1996, Carlberg+2014, Ceillier+2017}.  However, the fact that \tyc~is not rotating at a much larger fraction of its break-up speed disfavors a more equal-mass merger (e.g., two 1$M_{\odot}$ stars).

\textbf{Effective Temperature} 
Two spectroscopic and one photometric methods to determine the effective temperature, T$_{eff}$, of \tyc~were explored.  

\noindent
\textbf{I. Spectroscopic T$_{eff}$:} 
Here, we describe two spectroscopic techniques to derive T$_{eff}$. The first uses a set of 35 mostly unsaturated Fe I lines with good  gf values, taken from the NIST database, of class B (accuracy $<$10\% error) or better.  The Fe I lines chosen for \tyc~possesses a non-negligible covariance between microturbulent velocity ($\xi$) and T$_{eff}$, due to two saturated Fe I lines near an excitation potential of 1.0 eV (EW$\sim$100m\AA). Abundances were derived for each Fe~I line, based on measured EWs, using the LTE spectrum synthesis code, MOOG, with NLTE corrections\cite{Lind+2017}. After iterating on the microturbulent velocity parameter, by requiring a flat trend of iron abundance with EW, the input model atmosphere T$_{eff}$ was varied until there was no slope of iron abundance with line excitation potential (EP). 
Naturally, this minimized the Fe~I abundance residuals. In this absolute analysis, we find T$_{eff}$ = 5,800 K and $\xi$ = 1.5 km/s, with 1$\sigma$ T$_{eff}$ uncertainty = 27 K due to random errors in EW and atomic gf values of the Fe I line transitions, which are the product of the atomic transition oscillator strength and the statistical weight of the lower level, and 1$\sigma$ $\xi$
uncertainty = 0.06 km/s due to random errors. Unfortunately, the paucity of available unsaturated Fe~I lines with good gf values in \tyc's spectrum at low EP leads to a significant covariance between T$_{eff}$ and $\xi$, which increases the total 1$\sigma$ T$_{eff}$ uncertainty = 41 K using this technique.

The second technique employs line-by-line differential abundances\cite{Koch&McWilliam08,o'malley+2017}. Abundances for each Fe I line are derived relative to the same line in a standard star whose atmosphere parameters, including T$_{eff}$ and [Fe/H], are accurately known.  As in the absolute analysis, the model that best estimates T$_{eff}$ for the star provides a zero slope solution in a differential abundance versus excitation plot, relative to the standard. Ultimately, this is relative to the Sun, whose T$_{eff}$ is accurately known and not dependent on model atmospheres. By using this technique, the adopted gf values for each line cancel-out when the difference with the standard is computed.  Therefore, any measurable clean line in both target and standard star spectra can be used, even if no gf values are known. 
In this differential analysis we used 42 lines from the visual Keck/HIRES CCD spectrum of \tyc, 16 of which are in common with the absolute analysis (above).  We took line abundance differences relative to star Hip 66815\cite{o'malley+2017}, giving a best fit T$_{eff}$ of 5,900K, and a microturbulence velocity $\xi$ = 1.3 km/s. We adopt a T$_{eff}$ uncertainty of 51K, equal to the quadrature sum of the excitation
temperature uncertainties for \tyc~and our standard. We speculate that the 0.2 km/s microturbulent velocity difference between absolute and differential methods is likely due to errors in the chosen standard. 

\noindent
\textbf{I. Photometric T$_{eff}$:} 
We adopt color--T$_{eff}$ calibrations\cite{RamirezMelendez2005} to estimate photometric effective temperature of \tyc.  
Various photometry exists for \tyc, including Tycho B, V and 2MASS JHK magnitudes.  The Tycho photometry of \tyc~have relatively large measurement uncertainties, at 0.066 and 0.071 magnitudes in B and V, respectively. Combined with the 2MASS uncertainties near 0.02 mag. per band, the T$_{eff}$ uncertainties for B-V, V-J, and V-K are large, at 240K, 191K and 108K respectively. A further complication arises because at least part of the infrared flux observed from \tyc~is associated with circumstellar material (see ``Spectral Energy Distribution''). Particularly, the K-band flux is significantly affected by the circumstellar material excess.  In this way, the V-K color will be underestimate the stellar T$_{eff}$.  However, the J-band contamination by circumstellar emission appears negligible. 
The V-band magnitude of \tyc~can be derived from the existing SDSS ugriz photometry using available transformations\cite{Jester+2005}. 
The transformation of the SDSS photometry give a Johnson V-band magnitude for \tyc~of 11.15 $\pm$ 0.02 magnitudes, resulting in a V-J = 1.201; the de-reddened\cite{Winkler+1996} (V-J)$_0$ = 1.121 $\pm$ 0.028 mag. The color--T$_{eff}$ relation for (V-J)$_0$ = 1.121 and [Fe/H]=-0.90 (see below) is 5,855 K, with the color uncertainty of 0.028 mag corresponding to 76 K. 

Combining result from all three methods, we converge on T$_{eff}$ = 5,850 $\pm$ 50 K.

\textbf{Stellar Properties and Abundance Analysis} 
The model atmosphere abundance analysis follows the methods described in various studies\cite{McWilliamRick1994, McWilliam+95, Koch&McWilliam08, McWilliam+13}.  The analysis uses the MOOG spectrum synthesis routine\cite{Sneden73} and a grid of model atmospheres\cite{Castelli&Kurucz94}.  The stellar atmosphere were linearly interpolated to the desired atmospheric parameter values.  Elemental abundances and atmosphere parameters were determined iteratively.  For each line, the abundance was determined by finding the input element abundance required to match the observed and theoretical equivalent width (EW).  For this analysis, we focus on abundances of Fe and $\alpha$-elements (e.g. O, Mg, Si, S, Ca, and Ti) relative to solar. 

The LTE abundances of several elements were computed with a KURUCZ model\cite{Kurucz1979}, T$_{eff}$ = 5,850 $\pm$ 50 K, log(g) = 2.75 $\pm$ 0.15 cm/s$^2$, [M/H] = -0.90, $\xi$ = 1.50 km/s. The surface gravity log(g) is derived by demanding [Fe I/H] = [Fe II/H]. The microturbulent velocity parameter, $\xi$, was determined by requiring that derived Fe~I abundances be independent of equivalent width (EW). We find that $\xi$=1.5 km/s gives consistent iron abundances from the 35 optimal Fe~I lines chosen to derive stellar abundances. Uncertainties on the parameters were derived from the scatter in the abundances derived from the individual Fe lines and the sensitivity of log(g) to the temperature\cite{McWilliam+95}.

Using the above model prescription, \tyc~displays low metallicity ([Fe/H] = -0.90 dex), suggesting \tyc~is an old, metal-poor star. Alpha-element abundances (O, Mg, Si, S, Ca, and Ti) are all consistent with [$\alpha$/Fe] = +0.4 dex, similar to those in the Milky Way thick disk and halo, suggesting nucleosynthesis by core-collapse supernovae.  The kinematics of \tyc, based on Gaia data, indicate disk-like motion, so the old thick disk is favored: chemically and kinematically, \tyc~appears to be an old, thick disk star.

\textbf{Mass of \tyc} 
The spectroscopic gravity (log~$g$=2.75) and stellar radius (R=10.5 R$_{\odot}$) of \tyc~imply a stellar mass M=2.17 M$_{\odot}$. Taken at face value, this relatively high mass, and thus young age ($<$1 Gyr), are inconsistent with membership of the thick disk population, as indicated by the chemical composition, location and kinematics of \tyc.

However, a reasonable gravity decrease of $\sim$0.3 dex would put the mass of TYC-2597-735-1 near 1.1M$_{\odot}$, with a main sequence (MS) age consistent with an old star from the Galactic thick disk population.  Abundance analysis experiments of similar stars\cite{Koch&McWilliam08,o'malley+2017,McWilliam+2013} show that an abundance difference between Fe~I and Fe~II of $\sim$0.1 dex corresponds to a surface gravity difference of $\sim$0.3 dex.  For our analysis of TYC-2597-735-1, the 1-$\sigma$ RMS scatter of Fe~I and Fe~II abundances is 0.078 and 0.053 dex, respectively; but the 1$\sigma$ error on the mean difference is only 0.02 dex.  Thus, random error on the means of the abundances are unlikely to be the source of a 0.1 dex ionization equilibrium difference.  Systematic abundance errors, such as those due to gf value zero-points, model atmosphere effects, H-fraction, and other evolutionary effects may reduce log~$g$ to be more consistent with $\sim$1 M$_{\odot}$. Given the uncertainties in the mass estimate associated with using spectroscopically-derived stellar parameters, we constrain the mass of \tyc~between 1 -- 2.1 M$_{\odot}$ and expect the lower mass end to be more representative of the true mass of \tyc, given the strong evidence linking it to the Galactic thick disk population.

\subsection{MESA Models} 
We use the stellar evolution code MESA\cite{Paxton+19} to explore the impact of a stellar merger on the long-term stellar properties displayed by \tyc\cite{Metzger+17}. We assume a $2.17 M_{\odot}$ primary star and varied the mass of the merging secondary ($M_{\rm c}$) from those of Jupiter-mass planets up to low mass stars $M_{\rm c} \approx 0.3M_{\odot}$. We explore the merger evolution for the upper limit primary stellar mass inferred from our stellar properties analysis (see {\bf Mass of TYC 2597-735-1}), but we expect that, if we fix the stellar radius and structure of the primary and decreased the mass by a factor of 2 (to match the lower limit of the primary mass), expect to increase the companion mass by a factor of 2 (to match the energy). The lower mass primary star and higher mass companion, still safely within the low-mass star mass range, would yield roughly the same timescales of merger evolution. We expect that the luminosity of the lower-mass primary case would change in a non-trivial way, resulting in quantitative, but likely not qualitative, changes.  Future modeling efforts of the lower mass primary scenario is necessary to confirm this finding.

We consider different evolutionary stages of the primary at the time of the merger, ranging from it just leaving the main sequence, to an evolved sub-giant star on the horizontal branch. We simulate different evolutionary stages of the primary star by changing the radius of the star, ranging from 3 -- 10 R$_{\odot}$. For each simulation, right after the companion star plunges into the primary, we deposit energy into the envelope following the dynamical friction-driven inspiral. We  follow the star's subsequent evolution over a long timescale, $\sim$10,000 years. We then search for the combination of pre-merger parameters that best reproduce the present-day properties of \tyc, namely its luminosity, effective temperature, and surface gravity. We find where along in the merger's evolutionary track that the present-day properties of \tyc~best match its simulated progenitor model to independently estimate the time elapsed since the merger took place. The change in luminosity over time ($dM_{B}$/$dt$) is provided as a change in Johnson B-magnitude over the time of the merger and is used to independently verify the expected change in luminosity of \tyc~over a long time frame. 

ED Figure~\ref{fig:mesa} shows the evolution of the \texttt{MESA} models following the deposition of energy during the merger and assumes a $0.1 \, M_{\odot}$ companion. Different colored lines correspond to different sizes of the $2.17 \, M_{\odot}$ primary star at the time of merger, ranging from $3-10 \, R_{\odot}$.  We see that the moderately evolved sub-giant star with a radius of $5R_{\odot}$ does the best job at reproducing the present-day properties of \tyc~for an assumed time since merger of 1,000 years.  This age is a factor of $\sim 2$ smaller than that estimated for the BRN of $\gtrsim $2,000 years.  

We note that both the primary star's effective temperature and surface gravity 1,000 years after the merger energy injection adequately match the present-day derived values measured from optical spectroscopy and photometry of \tyc, which explains why \tyc's stellar properties are slightly skewed away from the bulk of moderately-evolved stars in the $T_{eff}$-log$g$ plane\cite{Afsar+18} (ED Figure~\ref{fig:tefflogg}). As demonstrated in ED Figure~\ref{fig:mesa}, the primary stellar properties continue to settle back towards equilibrium after the modeled merger takes place, its surface gravity increases, shifting \tyc's $T_{eff}$-log$g$ relationship in-line with other moderately-evolved stars around its effective temperature. 

Our simple one-dimensional models neglect a variety of effects, which if improved upon in future work would enable a more precise comparison to data.  These include, for example, the back reaction of the addition of mass (particularly unprocessed hydrogen) on the long-term stellar structure, as well as multi-D effects due to rotational mixing and the delayed accretion of mass from the remnant accretion disk.  More detailed models for angular momentum transport in the remnant will also be required to make specific predictions for the present-day rotation rate of \tyc.

\subsection{Long Term Light Curve of \tyc}
With the possibility of a merger millennia ago, it is reasonable to search for residual slow fading in \tyc's luminosity.  Slow evolutionary fading can only be detected with something like a century-long, well-calibrated light curve, and that essentially requires the use the photographic plates now archived at Harvard.  Examples include the sporadic fading of the Boyajian Star (KIC 8462852)\cite{Schaefer+16a,Schaefer+18}, brightening and fading of four ``Hot RCB stars''\cite{Schaefer+16b}, and the slow-then-fast fading of the `Stingray' planetary nebula nucleus\cite{Schaefer&Edwards+15}.

The photometric accuracy must be around 0.01 mag, for binned magnitudes, and both the old and modern magnitudes must be very carefully placed onto the identical magnitude scale.  Our procedure is to perform differential photometry of \tyc~with respect to the average of three nearby comparison stars of closely similar color and magnitude.  We chose TYC 2597-1026-1, TYC 2597-458-1, and TYC 2588-182-1, and adopted B magnitudes from the AAVSO Photometric All-Sky Survey (APASS) of 11.506, 11.513, and 11.912 mag respectively.  By using the same three comparison stars with similar color and only detectors with a sensitivity close to the Johnson B system, all color terms and systematic errors will be negligibly small, despite the many plates and detectors over the last century.  By averaging together many images from many nights, we can beat down measurement and systematic errors to usefully small values.
 
For \tyc, over two thousand plates from 1892--1989 exist.  The native spectral sensitivity of almost all the plates is effectively the original definition of the Johnson B magnitude system, and the comparison star magnitudes from APASS are accurately in the Johnson B band, so the resultant magnitudes are closely modern B magnitudes.  Fortunately, the Digital Access to a Sky Century $@$ Harvard (DASCH) program has already scanned all the relevant plates, and performed a good photometric analysis.  Critically, the DASCH calibration with the B-band APASS input catalog must be used, as the other calibrations lead to substantial systematic errors from imperfectly-corrected color terms.  The DASCH magnitudes were rejected for photometric uncertainty $>$0.30 mag, yellow and red sensitive plates, problem flag (AFLAG) values $>$50,000, plates where the target is within 0.30 mag of the measured plate limit, and $>$5-$\sigma$ outliers.  This leaves us with 2,077 B magnitudes.  These have no evidence of fast variations, so the light curve was averaged into 5-year bins.  The five-year bins from 1890--1895 and 1955--1970 were not used further because they have few plates.  The uncertainty in the binned magnitudes is the RMS scatter divided by the square root of the number of plates, with these uncertainties being consistent with a reduced chi-square of unity for a smooth curve.

To get a post-1989 light curve, the modern data must have a B-band spectral sensitivity, and a photometric error of around 0.01 mag or so.  Unfortunately, most modern data, either published or on-line, are not useable.  Gaia, ATLAS, ZTF, Pan-STARRS, ROTSE, and the Catalina Sky Survey have native photometric systems far from the B-band, so they cannot be reliably transformed to B-band with the needed accuracy.  Neither the literature, ASAS, nor the AAVSO database include our target star.  The Tycho BT magnitude has too large a quoted uncertainty ($\pm$0.07 mag) to be useful.  The only useable B magnitude that we can find is from APASS with two observations on 2012.2, for B equal to 11.756$\pm$0.004 mag.

Given the lack of pre-existing B magnitudes that can be accurately cross-calibrated with other detectors, we have commissioned CCD photometry.  Andrew Monson used the Three-hundred MilliMeter Telescope (TMMT\cite{Monson+2012}) in California to get 65 images on 22 nights from May 2014 to September 2015.  Lee McDonald used the 0.25-meter iTelescope 5 system in New Mexico to get 10 images on two nights in November 2019.  Kenneth Menzies used the 0.50-meter iTelescope 11 system in New Mexico to get 20 images on two nights in November 2019.  Ray Tomlin used an 8-inch f/6.3 Schmidt Cassegrain telescope in Illinois to get 25 images on two nights in November 2019.  All images were taken through a B filter, processed with the usual bias/dark/flat corrections, star magnitudes measured with aperture photometry, \tyc's magnitudes derived as differential from the same three comparison stars, the final magnitude taken as the average of all input images, and the formal uncertainty taken as the RMS scatter divided by the square root of the number of input images.  The RMS scatter between the magnitudes for the latter four observational runs is 0.008 mag, which is substantially larger than the formal error bars. \tyc~is likely not varying on any fast time scales, so 0.008 mag is a measure of how accurately photometry can be compared between observers with different systems, despite having the best possible conditions and procedures.  These observer-to-observer differences can be beaten down by averaging over the four observers, to get a modern 2015--2019 B magnitude of 11.767$\pm$0.004.

Our final light curve is presented in Supplemental Table~\ref{tab:lc} and ED Figure~\ref{fig:lc}.  We see a steady decline, from 1897.5 to 1952.5, consistent with a simple linear function with a slope of 0.127$\pm$0.016 mag/century (with a reduced-$\chi^2$ near unity).  This decline is significant at the 6-$\sigma$ level, as taken from an F-Test.  A linear fading of a light curve in magnitudes is the same as an exponential decline in the flux.  Between 1952.5 and 2019.9, the slope has flattened out and is consistent with a zero slope.  The two high points in the 1970s might represent some variability during the 1952.5--2019.9 interval, or maybe some residual errors of some sort, or perhaps a shallow decline from 1940 to 2019.9.  In all cases, \tyc~starts with a fast decline in the first half of the 1900s, and a slow-or-zero decline from 1950 to now.  A formal $\chi^2$ fit to a bent line gives a break around 1940, with slopes of 0.131$\pm$0.014 and 0.040$\pm$0.006 mag/century before and after the break.  An F-test shows the break to be significant at the 3.2-$\sigma$ level. The total B-mag decay is found to be between 0.11-0.12 mag from 1895-2020, consistent with $\sim$0.09 - 0.1 mag/century.

\subsection{Rates of BRN Formation} 
A fraction $f_{\rm bin} = 0.2-0.4$ of $\sim$ solar-mass stars have binary companions of mass $\gtrsim 0.1 M_{\odot}$\cite{Moe&DiStefano17}, of which a fraction $f_{\rm interact} \sim 0.1$ of these are on sufficiently short orbital separations $\lesssim 0.1-1$ AU to interact with the star due to tidal inspiral following its post-main sequence evolution\cite{Sato+08}.  Also assume that the current state of the system is observable for a time comparable to its present age, e.g. $t_{\rm obs} \sim 2t_{\rm age} \sim 10^{4}$ yr; this is motivated in part by the fact that in older sources the shock-heated electrons potentially responsible for exciting the $H_2$ would have adiabatically cooled by further expansion of the ejecta.  Given the lifetime of a 2$M_{\odot}$ star of $t_{\star} \sim 1$ Gyr, we then expect that only a fraction
\be
f_{\rm BRN} \approx f_{\rm bin}f_{\rm interact}\frac{t_{\rm obs}}{t_{\star}} \sim 5\times 10^{-8}.
\label{eq:fBRN1}
\ee
of A-stars to contain a BRN.  The number density of stars in the stellar neighborhood of mass $\gtrsim 2M_{\odot}$ is $n_{\star} \sim 5\times 10^{-4}$ pc$^{-3}$\cite{vanleeuwan2007}. Therefore, the total number of stars within a radius $R = D = 1.93$ kpc and scale-height $H \sim 1.5$ kpc is approximately $N_{\star} = n_{\star}\pi D^{2}H \sim 10^{7}$.  Thus, 
the number of candidate A-stars as close as \tyc~that we would expect to show a BRN is
\be
N_{\rm BRN} = N_{\star}f_{\rm BRN} \sim 0.4,
\ee
i.e.~of order unity. 

A second independent constraint on the rate comes from the Galactic rate of stellar merger events, as inferred from observations of luminous red nova transients, which Kochanek et al. 2014\cite{Kochanek+14} estimate to be $\mathcal{R} \sim 0.3-0.5 $ yr$^{-1}$.  We would therefore expect a total number $\tilde{N}_{\rm BRN} \sim \mathcal{R}t_{\rm obs} \sim 3,000$ Galactic A-stars in a BRN-bearing present state similar to \tyc~out of the total number $\tilde{N}_{\star} \sim 10^{9}$ of A-stars in the Milky Way (which we take to be a fraction $\sim 1\%$ of the total number $4\times 10^{11}$ of stars).  If every merger produced a BRN, then the expected BRN fraction would therefore be
\be
\tilde{f}_{\rm BRN} \sim \frac{\tilde{N}_{\rm BRN}}{\tilde{N}_{\star}} \sim 10^{-6},
\ee
which is consistent (within an order of magnitude) with the equally-uncertain estimate (\ref{eq:fBRN1}). The number of candidate A-stars from the rate of Galactic stellar mergers is then
\be
N_{\rm BRN} = N_{\star}f_{\rm BRN} \sim 10.
\ee
Therefore, we expect from our two independent methods to find anywhere between 0.4 - 10 BRNs. Our finding of one BRN is consistent with this rate.

\subsection{Analytic Processes Describing Mergers}
The process of the companion falling into the stellar envelope causes the ejection of mass in a rotationally supported disk (or decretion disk) near or above the surface of the star.  Mass ejected from the $L_{2}$ point during the earliest stages of a stellar merger can remain gravitationally bound\cite{Pejcha+16,Pejcha+16a}, particularly for binary mass ratios $q \equiv M_{\rm c}/M_{\star} \lesssim 0.06$, as would be satisfied in our case for $M_{\rm c} \lesssim 0.1M_{\odot}$.  The equatorial outflow acts to shape matter which is ejected at higher velocities during the final, plunge phase of the merger, into a bipolar outflow\cite{Metzger&Pejcha17,MacLeod+18}.  Our MESA calculations demonstrate that the merger process could eject a mass $M_{\rm ej} \sim 0.01M_{\odot}$ from the stellar envelope.

After forming a disk of radius $\sim R_{\star,0}$ and mass (of, say) $M_{\rm d,0} \sim  M_{\rm c} \sim 0.-1M_{\odot}$, the disk will accrete onto the star at a characteristic rate\cite{Metzger+12}
\be
\dot{M}_{\rm pk} \sim \frac{M_{\rm d,0}}{t_{\rm visc}} \sim 6\times 10^{25}{\rm g\,s^{-1}}\left(\frac{\alpha}{0.1}\right)\left(\frac{M_{\rm d,0}}{0.1M_{\odot}}\right)\left(\frac{M_{\star}}{M_{\odot}}\right)^{1/2}\left(\frac{R_{\rm d}}{3R_{\odot}}\right)^{-3/2}\left(\frac{H/R_{\rm d}}{0.5}\right)^{2},
\ee
over a timescale set by the viscosity of the disk
\be
t_{\rm visc} \sim 4\times 10^{5}{\rm s}\left(\frac{\alpha}{0.1}\right)^{-1}\left(\frac{M_{\star}}{M_{\odot}}\right)^{-1/2}\left(\frac{R_{\rm d}}{3R_{\odot}}\right)^{3/2}\left(\frac{H/R_{\rm d}}{0.5}\right)^{-2} \sim {\rm days},
\label{eq:tvisc}
\ee
where $\alpha$ is the Shakura-Sunyaev viscosity parameter.  We have assumed a thick disk with a scale-height $H \sim R_{\rm d}/2$, which is justified because the peak accretion luminosity
\be
L_{\rm acc,pk} = \frac{GM_{\star}\dot{M}_{\rm pk}}{R_{\star}} \sim 10^{41}{\rm erg\,s^{-1}}
\ee 
is several orders of magnitude larger than the Eddington luminosity of the star $\sim 10^{38}$ erg/s.

At such highly super-Eddington accretion rates, the disk cannot cool on the accretion timescale and is therefore subject to powerful outflows\cite{Narayan&Yi95}.  These outflows may carry away an order unity fraction of the initial disk mass in the form of a wind, i.e. $\sim 0.01M_{\odot}$.  This wind would also be funneled by the disk geometry into bipolar conical geometry, perhaps contributing to the BRN ejecta on top of the material ejected dynamically during the merger itself.  

What happens to the disk that remains bound to the star?  For a thick disk with constant $H/R_{\rm d}$ that evolves due to the viscous redistribution of angular momentum, the accretion rate onto the star evolves with time approximately as\cite{Pringle81,Metzger+08}
\be
\dot{M}(t) = \dot{M}_{\rm pk}\left(\frac{t}{t_{\rm visc}}\right)^{-4/3}, t \gg t_{\rm visc}
\ee
and therefore the disk will evolve to become sub-Eddington ($L_{\rm acc} \lesssim 2\times 10^{38}$ erg/s) after a timescale
\be
t_{\rm Edd} \sim 100t_{\rm visc} \sim {\rm 1\,\,year}
\ee
The current accretion rate would be estimated as
\be
\dot{M}(t_{\rm age}) =  \dot{M}_{\rm pk}\left(\frac{t_{\rm age}}{t_{\rm visc}}\right)^{-4/3} \sim 10^{19}{\rm g/s},
\ee
comparable to that inferred from the present-day H$\alpha$ luminosity (see {\bf Methods}).

As the disk accretes, it will also viscously spread outwards in radius due to the redistribution of angular momentum.  If the evolution of the disk conserves total angular momentum $J_{d} \propto M_{\rm d}(GM_{\star}R_{\rm d})^{1/2}$, where $M_{\rm d}$ and $R_{\rm d}$ are the total mass and outer disk radius, then the outer edge of the disk viscously spreads outwards in time as $R_{\rm d} \propto M_{\rm d}^{-2}$, i.e.
\be
R_{\rm d} = R_{\star}\left(\frac{t}{t_{\rm visc}}\right)^{2/3}
\ee
By the time the disk has become sub-Eddington ($t \sim t_{\rm Edd}$) its outer edge would reach a radius $R_{\rm d}(t_{\rm Edd}) \sim 1$ AU.  The disk may spread a bit further than this after $t_{\rm Edd}$, but since the disk will become geometrically thin ($H/R_{\rm d} \ll 1$) after becoming sub-Eddington the viscous timescale over which the spreading occurs $t_{\rm visc} \propto (H/R_{\rm d})^{-2}$ (eq.~\ref{eq:tvisc}) will increase significantly.  Thus, we expect $R_{\rm d}(t_{\rm age}) \sim$ few 10$^{13}$ cm.  The equilibrium temperature at the outer disk edge
\be
T_{\rm eq} = \left(\frac{L_{\star}}{4\pi \sigma R_{\rm d}^{2}}\right)^{1/4} \sim 500{\rm K},
\ee
i.e. sufficiently low to allow dust formation and consistent with the inferred temperatures of the NIR excess of \tyc~(see {\bf Spectral Energy Distribution}).  

The current mass of the disk in such a scenario is approximately $M_{\rm d,0}(t_{\rm Edd}/t_{\rm visc})^{-1/3} \sim 0.1M_{\rm d,0} \sim 10^{-3}M_{\odot}$, which, assuming the standard $1:100$ value for the dust-gas mass ratio, would imply a present dust mass of $\sim 10^{-5}M_{\odot}$. This is consistent with the minimum mass of dust \emph{only} $\gtrsim 5\times 10^{-9}M_{\odot}$ we derive from model-fitting the IR excess\cite{Jura1998} (see {\bf Spectral Energy Distribution}). 


\newpage

\bibliography{refs}

\newpage

\renewcommand{\tablename}{Supplemental Table}

\begin{table}
\centering
\small
\caption{\tyc~light curve.} 
\label{tab:lc}
\begin{tabular}{lll}
\hline
Source	& $\langle$Year$\rangle$    & 	$\langle$B$_{BRN}\rangle$ (mag)	   \\
\hline
DASCH   &	1897.5   &	11.660 $\pm$ 0.023\\
DASCH   &	1902.5   &	11.675 $\pm$ 0.014\\
DASCH   &	1907.5   &	11.708 $\pm$ 0.015\\
DASCH   &	1912.5   &	11.719 $\pm$ 0.023\\
DASCH   &	1917.5   &	11.709 $\pm$ 0.016\\
DASCH   &	1922.5   &	11.722 $\pm$ 0.023\\
DASCH   &	1927.5   &	11.709 $\pm$ 0.015\\
DASCH   &	1932.5   &	11.726 $\pm$ 0.010\\
DASCH   &	1937.5   &	11.734 $\pm$ 0.007\\
DASCH   &	1942.5   &	11.737 $\pm$ 0.007\\
DASCH   &	1947.5   &	11.724 $\pm$ 0.008\\
DASCH   &	1952.5   &	11.764 $\pm$ 0.012\\
DASCH   &	1972.5   &	11.731 $\pm$ 0.016\\
DASCH   &	1977.5   &	11.747 $\pm$ 0.011\\
DASCH   &	1982.5   &	11.759 $\pm$ 0.012\\
DASCH   &	1987.5   &	11.761 $\pm$ 0.011\\
APASS   &	2012.5   &	11.756 $\pm$ 0.004\\
TMMT    &	2015.5   &	11.764 $\pm$ 0.001\\
AAVSO   &	2019.9   &	11.780 $\pm$ 0.003\\
AAVSO   &	2019.9   &	11.758 $\pm$ 0.003\\
AAVSO   &	2019.9   &	11.764 $\pm$ 0.003
\\
\hline
\end{tabular}
\end{table}


\clearpage

\end{document}